\documentclass[11pt, a4paper]{article}

\usepackage{booktabs}
\usepackage[toc,page]{appendix}
\usepackage{amsmath}
\usepackage{amsfonts}
\usepackage{amssymb}
\usepackage{graphicx, rotating}
\usepackage{epstopdf}
\usepackage{epsfig}
\usepackage{braket}
\usepackage{simplewick}
\usepackage{bbm}
\usepackage{empheq}
\usepackage{latexsym}
\usepackage{graphicx}
\usepackage{leftidx}
\usepackage{subcaption}
\usepackage{color}
\usepackage{amsmath,amssymb}
\usepackage{cite}
\usepackage{slashed}
\usepackage{hyperref}
\usepackage{xcolor}
\hypersetup{colorlinks, citecolor=bluscuro, linkcolor=bluscuro, urlcolor=bluscuro} 
\definecolor{rossos}{cmyk}{0,1,1,0.55}
\definecolor{bluscuro}{rgb}{0.15, 0.2, .85}
\definecolor{bluchiaro}{cmyk}{1,.3,0.,0.1}

\usepackage{multirow}
\usepackage[hmarginratio=1:1,top=32mm,columnsep=20pt]{geometry}

\numberwithin{equation}{section}

\newcommand{\fref}[1]{Fig.~\ref{#1}} 
\newcommand{\eref}[1]{Eq.~\eqref{#1}}

\newcommand{\aref}[1]{App.~\ref{#1}}
\newcommand{\sref}[1]{Sec.~\ref{#1}}
\newcommand{\cref}[1]{Chapter~\ref{#1}}
\newcommand{\tref}[1]{Tab.~\ref{#1}}

\newcommand{\nnl}{\nonumber \\}

\newcommand{\beq}{\begin{equation}} 
\newcommand{\eeq}{\end{equation}} 
\newcommand{\ba}{\begin{array}}  
\newcommand{\ea}{\end{array}} 
\newcommand{\bea}{\begin{eqnarray}}  
\newcommand{\eea}{\end{eqnarray} }  
\newcommand{\be}{\begin{eqnarray}}  
\newcommand{\ee}{\end{eqnarray} }  
\newcommand{\bal}{\begin{align}}
\newcommand{\eal}{\end{align}}   
\newcommand{\bi}{\begin{itemize}}  
\newcommand{\ei}{\end{itemize}}  
\newcommand{\ben}{\begin{enumerate}}  
\newcommand{\een}{\end{enumerate}}  
\newcommand{\bc}{\begin{center}}
\newcommand{\ec}{\end{center}} 
\newcommand{\bt}{\begin{table}}
\newcommand{\et}{\end{table}}  
\newcommand{\btb}{\begin{tabular}}
\newcommand{\etb}{\end{tabular}}

\newcommand{\cM}{{\mathcal M}}

\newcommand{\mpl}{M_{\mathrm Pl}}

\def\mpl{\, M_{\rm Pl}}

\def\ra{\rangle}
\def\la{\langle}  

\newcommand{\im}{{\mathrm{Im}} \,}

\newcommand{\eps}{\epsilon}

\newcommand{\5}{{\mathbf 5}}
\newcommand{\6}{{\mathbf 6}}
\newcommand{\p}{{\mathbf p}}

 \def\mpl{m_\mathrm{Pl}}
 \def\Disc{\mathrm{Disc}\,}

\begin{document}
\begin{center}

\vspace*{-25mm}

\begin{flushright}
{\small Saclay-t21/003}\\
CERN-TH-2021-029\\
NUHEP-TH/21-07\\
\end{flushright}

\vspace{2cm}
{\Large \bf  Gravitational Causality and the Self-Stress of Photons}
\vspace{1.4cm}\\
{Brando Bellazzini$\,^{1,2,3}$, Giulia Isabella$\,^{1,4}$, \\
Matthew Lewandowski$\,^{5}$, and Francesco Sgarlata$\,^{6}$}

 \vspace*{.5cm} 
\begin{footnotesize}
\begin{it}
 $^1$ Universit\'e Paris-Saclay, CNRS, CEA, Institut de Physique Th\'eorique, 91191, Gif-sur-Yvette, France.  \\
 $^2$ CERN, Theoretical Physics Department, Geneva, Switzerland\\
 $^3$Theoretical Particle Physics Laboratory (LPTP), Institute of Physics, EPFL, Lausanne, Switzerland \\
$^4$ Universit\'e Paris-Saclay, CNRS/IN2P3, IJCLab, 91405 Orsay, France\\
$^5$ Department of Physics and Astronomy, Northwestern University, Evanston, IL 60208, USA \\
$^6$ Department of Physics, LEPP, Cornell University, Ithaca, NY 14853, USA 

\end{it}
\end{footnotesize}

\vspace*{.2cm} 

\vspace*{10mm} 
\begin{abstract}\noindent\normalsize
We study causality in gravitational systems beyond the classical limit.
Using on-shell methods, we consider the 1-loop corrections  from charged particles to the photon energy-momentum tensor ---the self-stress--- that controls the quantum interaction between two on-shell photons and one off-shell graviton. The self-stress determines in turn the phase shift and time delay in the scattering of photons against a spectator particle of any spin in the eikonal regime. 
We show that the sign of the $\beta$-function associated to the running gauge coupling  is related to the sign of time delay at small impact parameter. 
Our results show that, {at first post-Minkowskian order}, asymptotic causality, where the time delay experienced by any particle must be positive, is respected quantum mechanically. 
Contrasted with asymptotic causality, we explore a local notion of causality, where the time delay is longer than the one of gravitons, which is seemingly violated by quantum effects.

\end{abstract}
\end{center}
\newpage
{ \hypersetup{hidelinks} \tableofcontents }
\newpage

\section*{Introduction}

Causality is a cornerstone of relativistic quantum field theory (QFT), with one of its most profound implications being the existence of anti-particles.   
Furthermore, causality has important implications for properties of scattering amplitudes in flat space, such as analyticity in the complex plane of Mandelstam variables. In combination with  unitarity, causality enforces non-trivial consistency conditions on effective field theories (EFTs) that emerge at low-energy from underlying causal and unitary QFTs, often in the form of ``positivity constraints'' on the EFT's Wilson coefficients that enter in 4-point scattering,  see e.g. \cite{Adams:2006sv}.

The notion of causality in the presence of gravity is certainly more subtle because the spacetime metric that defines the causal structure is itself subject to quantum fluctuations. Moreover, quantum fluctuations give rise to different light-cones for the various species of particles. 

A fundamental step in understanding the role of causality in gravity has been taken in \cite{Camanho:2014apa}, where the properties of 3-point vertices involving at least one graviton have been linked to the tree level classical corrections of the time delay that particles experience in eikonal scattering. Requiring positive time delay over all range of impact parameters provides thus non-trivial causality constraints on the 3-point functions.

In this work we are interested in gravitational causality beyond the classical limit and study  the first non-trivial quantum effects. 
The question that we have in mind is the following: what notion of causality is respected ---quantum-mechanically--- once gravity generates spacetime backgrounds?   
When quantum loops are taken into account, is the theory causal with respect to a lightcone defined by graviton propagation (bulk causality), or rather with respect to the asymptotic Minkowski metric (asymptotic causality) in the vacuum?

We address these questions by studying eikonal scattering around flat spacetime perturbatively, where some spectator source weakly perturbs Minkowski space and generates a non-trivial scattering phase shift, hence a time delay or advance, for photons that are sent through such space.   We focus in particular on the gauge 1-loop corrections $o(g^2/16\pi^2)$  to the time delay, while working to the lowest post-Minkowskian order $o(1/\mpl^2)$, i.e. neglecting gravitational loop contributions.  

\begin{figure}[tb]
   \centering
  \includegraphics[width=0.35 \textwidth]{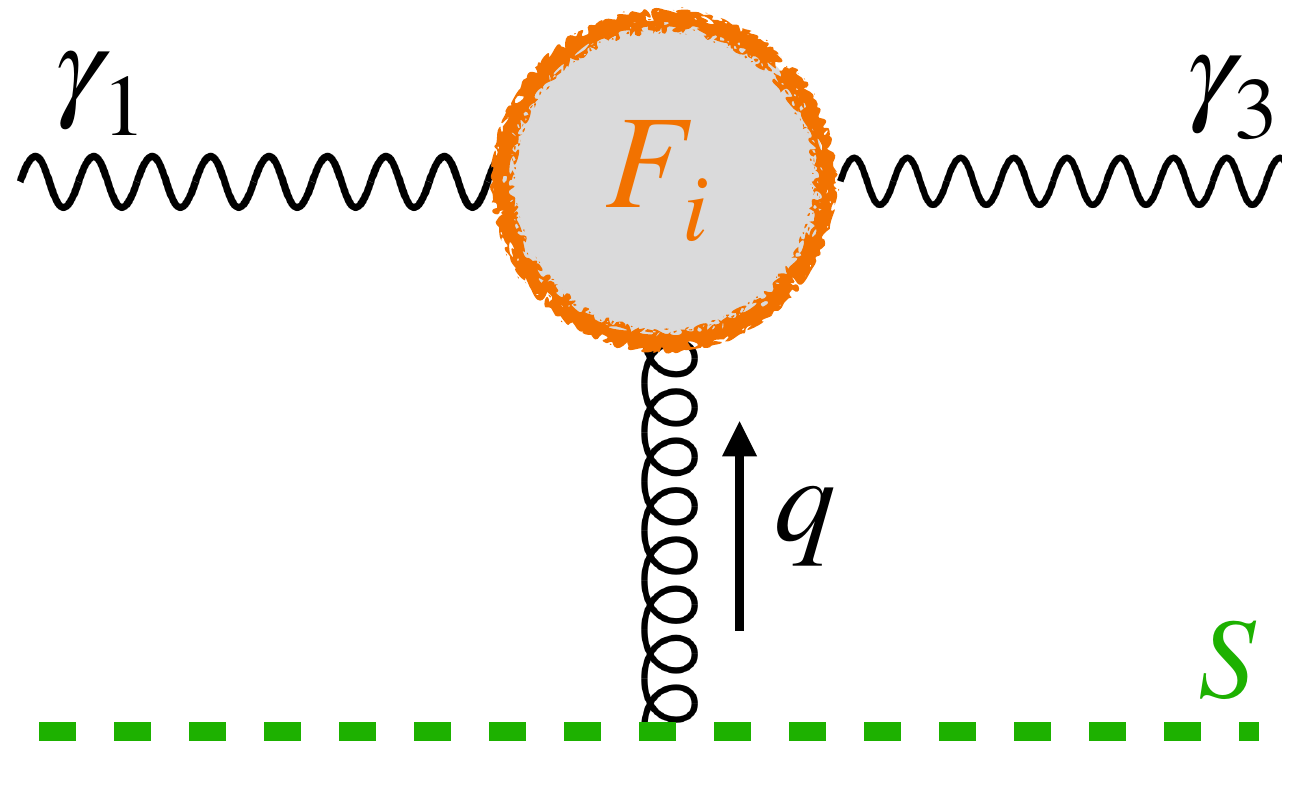}
  
\caption{  \label{Fi} \footnotesize
Type of diagram contributing to the eikonal scattering and the resulting time delay via the form factors $F_i$.  Curly lines are   graviton legs,  wiggle lines represent photons, dashed lines are the spectators, and $F_i$ are the form factors defined in \eref{formfactorsAll} associated to the photon energy-momentum tensor.}
\end{figure}

The causal response of photons in a perturbed Minkowski space is extracted by calculating the self-stress (energy-momentum tensor) of photon pairs at one loop. We consider loops of charged states of massive scalars, fermions, and vectors, which is equivalent to the full 1-loop correction to the 3-point function as predicted by the Standard Model.    The self-stress is  parametrized by three gravitational form factors $F_i(q^2)$, for $i = 1,2,3$ with $q$ the momentum of the exchanged graviton, that correspond to 3-point functions with off-shell gravitons having a non-trivial momentum dependence that in turn affects the time delay in the eikonal scattering, see \fref{Fi}.

The detailed computation of the form factors is performed via on-shell methods (unitarity cuts, massive spinor-helicity formalism, and dispersion relations), which are computationally powerful and conceptually  neat, avoiding the need to deal  with gauge-dependent quantities to extract physical observables. 
For wavelengths of the exchanged graviton  larger than the charged-particle Compton wavelength $1/m$ in the loop, we recover the classic results of  \cite{Drummond:1979pp} (extended to include spin-1 loops), from which the study of causality in the low-energy limit of quantum electrodynamics (QED) coupled to gravity originally started. At shorter graviton wavelengths, virtual particles can probe larger regions of spacetime.  As  a result, we show that the sign of the time delay at small impact parameters $b\ll 1/m$ is related to the sign of the QED $\beta$-function contribution from charged particles.  

We find no asymptotic-causality violation for impact parameters larger than the length scale associated to the Landau pole (below which our calculations no longer apply) in spinorial and scalar QED. Loops of spin-1 $W$-bosons do not generate a Landau pole\footnote{We are including Higgs bosons to make the theory renormalizable in the absence of gravity.} and give in fact no asymptotic-causality violation because of Sudakov infrared (IR) divergences which exponentiate and suppress the form factor at large momentum transfer. 
Instead, and despite being classically valid, we find that bulk causality is not respected quantum mechanically, within our setup.  

The remainder of this paper is organized as follows. In \sref{Tmunu} we calculate the energy-momentum tensor at one loop in the Standard Model and study its properties, including the connection between the gravitational form factors, $\beta$-functions, and IR divergences. We also study the Higgs/graviton mixing that contributes to the form factors. 
In \sref{Phaseshiftsection}, we calculate the phase shift by taking the eikonal limit of the amplitudes in the relevant kinematic configuration and computing its Fourier transform to impact parameter space. Different limits of the integration are studied analytically at large and small impact parameter.   \sref{causality} is devoted to studying the implications of the 1-loop self-stress on the two notions of causality. 
Conclusions and future directions are discussed in \sref{conclusion}. 

\section{The Photon Self-Stress}\label{Tmunu}
In this section we calculate the matrix element of a symmetric and conserved energy-momentum tensor $T_{\mu\nu}$ in flat spacetime 
 \begin{equation}
 \label{Tmatrix1}
 \langle 0| T_{\mu\nu}(x) | k^\prime  k \rangle = e^{-i(k+k^\prime) \cdot x}\langle 0| T_{\mu\nu}(0) | k^\prime  k \rangle   \,, \qquad T_{\mu\nu}=T_{\nu\mu}\qquad \partial_\mu T^{\mu\nu}=0
 \end{equation}
between a pair of (identical) incoming massless spin-1 particle states,  both taken on-shell, 
\begin{equation}
\label{onshellcond}
k^2=k^{\prime\,2}=0\,,\qquad \epsilon \cdot  k =\epsilon^\prime \cdot  k^{\prime}=0\, ,
\end{equation} 
where the dot $\cdot $ indicates Lorentz contraction with the Minkowski metric (see \aref{conventions} for conventions), and $k^2 \equiv k \cdot k$.  The $\epsilon$ and $\epsilon^\prime$ are polarization vectors associated, up to a gauge choice, to $k$ and $k^\prime$ respectively.\footnote{The little-group index that labels the helicity of the particles is sometime left understood to avoid clutter of notation, but displayed whenever relevant.} In analogy with low-energy quantum electrodynamics, we refer hereafter to these states as ``photons,'' although our analysis goes beyond real world QED to any massless spin-1 minimally coupled to gravity.   By crossing symmetry, \eref{Tmatrix1} determines as well the $\langle k^\prime  | T_{\mu\nu}(x) | k \rangle$ matrix element 
 by the replacement  $\epsilon^\prime \rightarrow \epsilon^{\prime *}$ and $k^\prime\rightarrow -k^\prime  $ in \eref{formfactorsAll}, which flips the helicity.

After Fourier transforming (\ref{Tmatrix1}) and factoring out a  $(2\pi)^4\delta^4(k+k^\prime+q)$ from momentum conservation, the matrix element can be written as the sum of three conserved and gauge-invariant tensor structures multiplied by scalar form factors $F_{i}(t)$, for $i = 1,2,3$, 
\begin{align}
\begin{split}  \label{formfactorsAll}
\langle 0 | T_{\mu\nu}(0) | k^\prime k \rangle \mathcal{N}= & \langle 0 | T_{\mu\nu}(0) | k^\prime k \rangle\big|^{\mathrm{tree}} \mathcal{N} \,F_1(t)  \\ 
& +  P_{\mu\nu} ( q ) \left[2(\epsilon^\prime \cdot k)(\epsilon \cdot  k^\prime)- q^2 (\epsilon \cdot \epsilon^\prime)\right] F_2(t) \\ 
& + p_\mu p_\nu \left[2(\epsilon^\prime \cdot k)(\epsilon \cdot k^\prime)- q^2 (\epsilon \cdot \epsilon^\prime)\right] F_3(t) 
\end{split}
\end{align}
where  we have defined \begin{equation}
p\equiv k^\prime-k\,, \qquad q=-(k+k^\prime)\,, \qquad  t=q^2 = 2 k \cdot  k^\prime  \,,\qquad P_{\mu\nu}(q)=q_\mu q_\nu-\eta_{\mu\nu} q^2 
\end{equation}
and $\mathcal{N}=\sqrt{4 |k^0 k^{\prime 0}|}$ is the relativistic normalization factor. 
The basis of tensor structures is chosen to isolate first the classical term
 \begin{equation}
 \langle 0 | T_{\mu\nu}(0) | k^\prime k \rangle\big|^{\mathrm{tree}}\mathcal{N}=\left(k_{[\mu} \epsilon_{\alpha]}k^{\prime}_{[\nu} \epsilon^{\prime}_{\beta]}+ k_{[\nu} \epsilon_{\alpha]}k^{\prime}_{[\mu} \epsilon^{\prime}_{\beta]} \right)\eta^{\alpha\beta}
 -\frac{1}{2}\eta_{\mu\nu}k ^{[\alpha}\epsilon ^{\beta]} k^\prime_{[\alpha}\epsilon^{\prime}_{\beta]}
 \end{equation}
associated to the free-photon $T^{(\gamma)}_{\mu\nu}$  in \eref{photonTmunu}, then the identically conserved terms $P_{\mu\nu}(q)$ associated to the so-called improvement terms, and finally  the projector $p_\mu p_\nu$ which is orthogonal to $q_\mu$, and hence conserved,  via the on-shell condition. Their physical meaning is made manifest by the dependence on the helicities $h$ and $h^\prime$  
  \begin{equation}
  \label{MatrixTmunu}
\!\!\!\!\!\! \langle 0|T^{\mu\nu}(0)| k^{\prime h^\prime} k^h \rangle \mathcal{N} =  \left(
\begin{array}{ll}
{1\over 2} \langle k^\prime \sigma^\mu k]\langle k^\prime \sigma^\nu k]F_1(t) &-\langle k k^\prime \rangle^2 \left(     P^{\mu\nu}(q) F_2(t)+p^\mu p^\nu F_3(t)  \right) \\
-[k k^\prime ]^2\left(  P^{\mu\nu}(q)F_2(t)+p^\mu p^\nu F_3(t)\right)&  {1\over 2}\langle k\sigma^\mu k^\prime ]\langle k\sigma^\nu k^\prime]F_1(t) 
\end{array}
\right)
 \end{equation}
where the diagonal entries correspond to $h^\prime=-h=\pm $ (here referred to as helicity-preserving, in reference to the crossed process),  while the off-diagonal entries correspond to $h=h^\prime=\pm $ (helicity-flipping).  {Here, $\sigma^\nu$ are the Pauli matrices, and the square and angle brackets are the standard spinor helicity variables (see \aref{conventions}).}  
One can recognize the three covariant little-group structures:  $F_1$ parametrizes the helicity-preserving scattering against an off-shell graviton --equivalently on-shell massive spin-2---,    while $F_{2}$ and $F_{3}$ control the overlap between the helicity-flipping photon pair ---hence having zero spin in the direction of motion---  and either the spin-0 or the spin-2 state found in $T_{\mu\nu}|0\rangle$, which can have such a vanishing spin projection.  There is no spin-1 state and only one spin-0 state because of the conservation equation  $\partial_\mu T^{\mu\nu}=0$. 

From the normalization $\lim_{k^\prime\rightarrow k} \langle k^{\prime h} | T^{\mu\nu}(0) | k^h\rangle =  k^\mu k^\nu/ k^0$  associated to the particle 4-momentum $P^\mu|k\rangle=\int d^3x  \, T^{0\mu}(x)|k\rangle=k^\mu|k\rangle$, the helicity-preserving entries of \eref{MatrixTmunu} are fixed at zero-momentum transfer 
\begin{equation}
\label{normalizationF1}
F_1(t\rightarrow 0)=1\,.
\end{equation}
Once coupled to gravity,  this corresponds to the universal helicity-preserving low-energy coupling  of gravity set by the reduced Planck mass $\mpl=(8\pi G)^{-1/2}$, where $G$ is the Newton constant.

\subsection{Self-Stress at One Loop}\label{oneloopdisc}

The energy-momentum tensor we consider is defined operationally as the covariant source of a weak gravitational field.  At tree-level $F_1=1$ and $F_{2,3}=0$ for all values of $t$, corresponding to the photon matrix elements of the free $T_{\mu\nu}$ reported in \aref{FreeTmunu}.   At 1-loop, radiative corrections modify these values via loops of charged states coupled to the photons,\footnote{There are in general also loop corrections to $F_2$ from the vacuum expectation value (VEV) of \emph{neutral} scalars non-minimally coupled to gravity, $\propto \xi \int \sqrt{|g|} R H^2 +\ldots$, so that a non-vanishing $\langle H\rangle=v$ generates graviton/scalar mixing $\propto v\xi$. See Section~\ref{nonminSec} and Appendix~\ref{HiggsGravitymix}. } and in the following we reconstruct the radiative self-stress matrix elements from tree-level amplitudes using on-shell methods.  

\begin{figure}[tb]
\centering
\begin{minipage}{.33\textwidth}
  \centering

 \includegraphics[width=0.8 \textwidth]{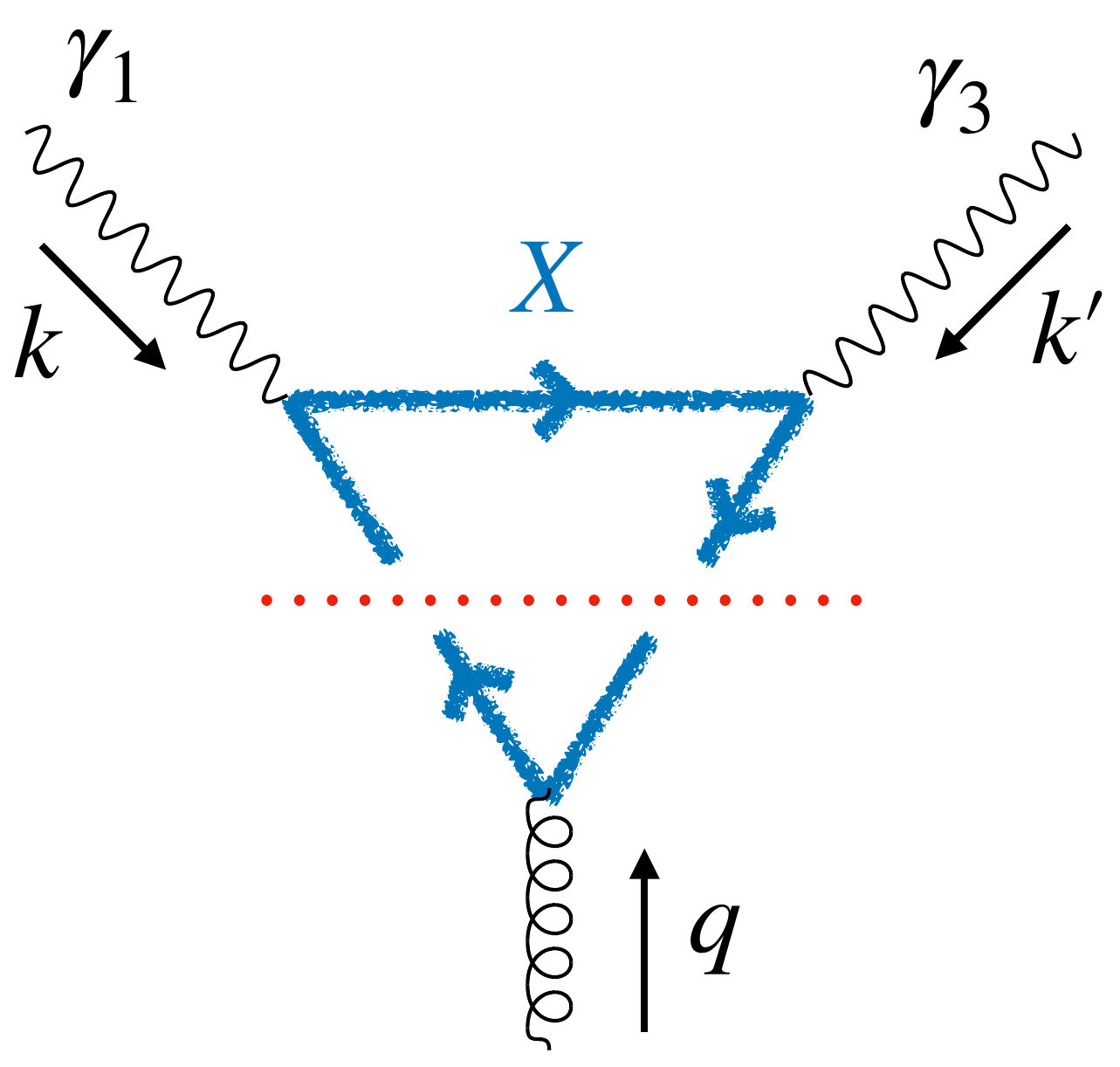}
   
\end{minipage}
\begin{minipage}{.33\textwidth}
  \centering
   
  \includegraphics[width=0.52 \textwidth]{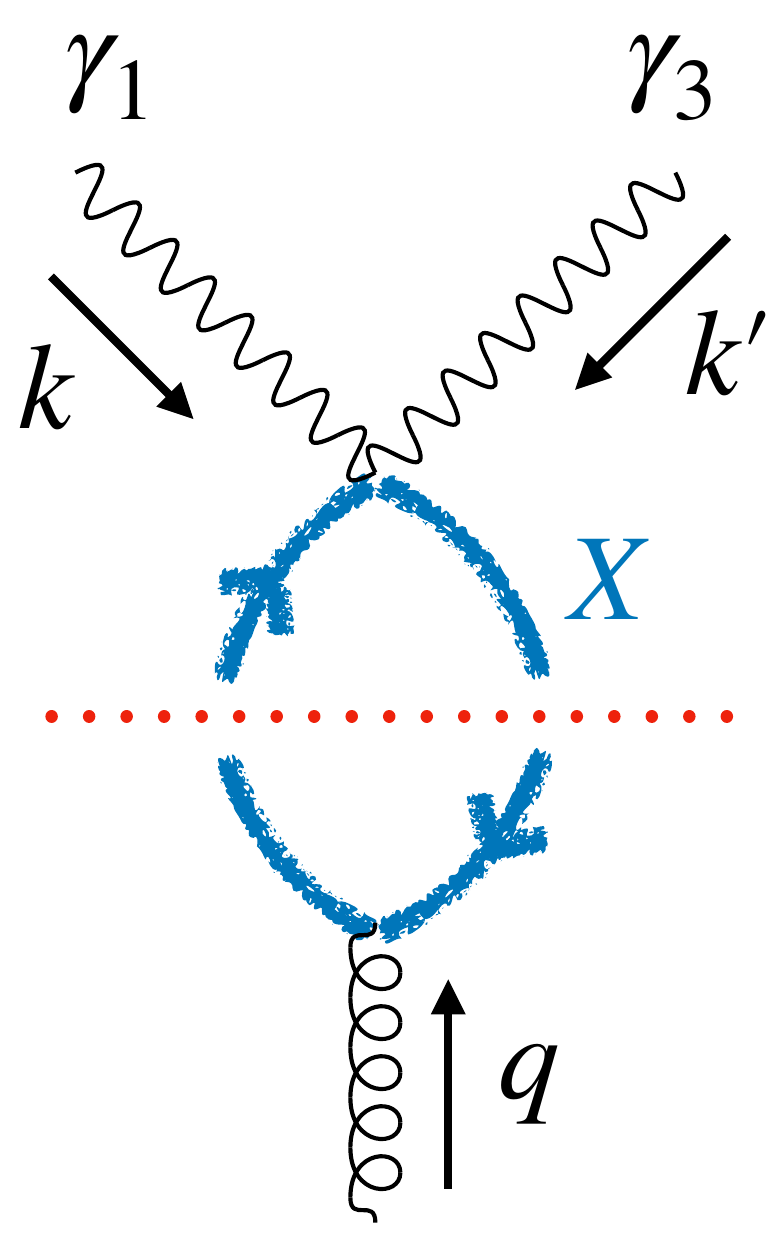}
  
\end{minipage}%

\caption{ \footnotesize \label{disc}
Diagrams contributing to the 1-loop discontinuity of the 3-point function with $k^2=k^{\prime\,2}=0$ and $q^2>4m_X^2$ (1-3 crossed triangle diagrams omitted for simplicity). Curly lines are graviton legs,  wiggle lines represent photons, charged particles of spin 0, 1/2 and 1 in the loop are represented by $X=\phi,\psi,W$ respectively, and dotted lines put  legs that they cut on-shell.
}
\end{figure}

One simple and efficient way to extract the form factors $F_i$ is calculating first their discontinuities in the complex $t$-plane across the real line for $t>4m^2$, as shown in the loop diagrams in {\fref{disc}},  where $m$ is the mass of any given charged state running in the loop. Then one computes the real parts by a simple dispersive integral, see \eref{discAllFs}. 
It turns out, in fact,  that the gravitational phase shift and the associated light-bending  and time delays can be extracted directly from the discontinuity alone (see for example Eqs.~\eqref{eigenvaluesPhaseShift}, \eqref{Fihat} combined with \eqref{polediscv3}).

The discontinuity at one loop can be calculated by either explicit evaluation of the (non-analytic part of the) triangle and bubble diagrams in {\fref{disc}} (with no cuts),  or equivalently by convoluting tree-level amplitudes via the Cutkosky rule. {We follow the latter approach and have found it} convenient to build first  an auxiliary 2-to-2 scattering amplitude $1_\gamma 3_\gamma\rightarrow 2_S 4_S$  for photons into some spectators $S$  taken to be a real massless scalar minimally coupled to gravity. The discontinuity  of the energy-momentum tensor in the Mandelstam variable $s_{13} \equiv (k_1+k_3)^2=t$ for $s_{13}>4m^2$ is promptly obtained from the auxiliary amplitude multiplied by $s_{13}\mpl^2$ 
\begin{equation}
\label{DiscForm0}
k_2^{(\mu} k_4^{\nu)} \Disc \langle 0|  T_{\mu\nu}(0) | k^{h_1}_1 k^{h_3}_3\rangle \mathcal{N} =\mpl^2 \Disc s_{13} \cM (1_\gamma 3_\gamma \rightarrow 2_S 4_S) 
\end{equation}
 by factoring out  $k_2^{(\mu} k_4^{\nu)}$. This is effectively the same as considering the $s_{13}$-channel discontinuity of 2-to-1 amplitudes associated to pairs of photons producing an off-shell graviton.  The right-hand side of \eref{DiscForm0} can be calculated at one loop via the Cutkosky rule
\begin{equation}
\label{cutkoskirule}
\Disc\cM(1_\gamma 3_\gamma \rightarrow 2_S 4_S)= i \int d\Pi_{56} \cM(1_\gamma 3_\gamma \rightarrow 5_X 6_{\bar X}) \cM(5_X 6_{
\bar X}\rightarrow 2_S 4_S)
\end{equation}
 using the tree-level amplitudes $\cM(1_\gamma 3_\gamma\rightarrow 5_X 6_{\bar X})$ and $\cM(5_X 6_{\bar X }\rightarrow 2_S 4_S)$ where $(5_X,6_{\bar X})$ is any pair of charged particles/antiparticles  of spin $J_X=0$, $1/2$ or $1$ in the Standard Model (hereafter dubbed $\phi$, $\psi$ and $W$ respectively), $d\Pi_{56}$ is their Lorentz invariant two-body phase space, and the sum over the helicities of internal particles is left understood. All the relevant amplitudes are summarized in \tref{amp}, and the diagrams contributing to the discontinuity are shown in \fref{disc}.
 
For $X=\phi,\ \psi$, the 4-point functions $\cM(1_\gamma 3_\gamma \rightarrow 5_X 6_{\bar X})$ are the pair production amplitudes in standard (scalar and spinorial) QED. They can either be obtained by Feynman diagrams from the Lagrangians given in \aref{lagrangianInt}, or  recovered from standard on-shell techniques. With the latter approach,  unitarity dictates the factorization of the 4-point amplitude into 3-point amplitudes which are completely fixed by little group scaling and dimensional analysis, (for reviews see e.g. \cite{Cheung:2017pzi,Elvang:2013cua}). 

The case of the massive vector $X=W$ is slightly more delicate because the high energy limits involve extra 3-point vertices relative to the one of massless Yang-Mills, reflecting the presence of the {\it eaten} Goldstone bosons. The minimal cubic coupling we consider is thus fixed by its high energy behavior, requiring that the vertices match massless Yang-Mills for the transverse polarizations,  and minimally coupled massless scalars for the longitudinal polarizations. This is simply the on-shell amplitude description of the Higgs mechanism \cite{Arkani-Hamed:2017jhn}, i.e. the Goldstone equivalence theorem.  Once again, this result is matched by the Lagrangian formulation of \aref{lagrangianInt}.  The last column of  \tref{amp} is the production of the neutral spectator through the gravitational interaction. All $X$ are taken to couple minimally to gravity except for the  non-minimal coupling present on \tref{amp} for $\phi$, parametrized by $\xi_\phi$. Such a contribution is discussed in \sref{nonminSec}. 

 \!\!\! \begin{table}[t]
    \centering
  \begin{tabular}{l|c|c|c}   \toprule 
& $ \cM(1_\gamma^- 3_\gamma^+ 5_X 6_{\bar X})$ & $ \cM(1_\gamma^- 3_\gamma^- 5_X 6_{\bar X})$ &$ \cM(5_X 6_{
\bar X} 2_S 4_S)$\\	\midrule
$\phi$ &${g^2\la1(k_5-k_6)3]^2\over  2\left(s_{15}-m^2\right)\left(s_{16}-m^2\right)}$ & $ {2 g^2m^2\la 13\ra^2\over \left(s_{15}-m^2\right)\left(s_{16}-m^2\right)}$& ${\left(s_{25}-m^2\right)\left(s_{45}-m^2\right)\over  \mpl^2 s_{24}}-\xi_\phi  {s_{24}\over 6 \mpl^2}$ \\[0.3cm] 
$\psi$ &${g^2\la1(k_6-k_5)3]\over (s_{15}-m^2)(s_{16}-m^2)} \left(\la 1\5 \ra[3\6]+\la 1\6 \ra[3\5]\right)$ & ${2 g^2m \la 13\ra^2\la\6 \5 \ra\over \left(s_{15}-m^2\right)\left(s_{16}-m^2\right)}$ & ${s_{25}-s_{45}\over 4 \mpl^2s_{24}}\left(\la \6(k_2-k_4)\5]+\la \5(k_2-k_4)\6]\right)$ \\[0.3cm] 
$W$ &${2g^2\over (s_{15}-m^2)(s_{16}-m^2)}\left(\la 1\5 \ra[3\6]+\la 1\6 \ra[3\5]\right)^2$ & ${2 g^2 \la 13\ra^2\la\6 \5 \ra^2\over \left(s_{15}-m^2\right)\left(s_{16}-m^2\right)}$ & ${-1\over 4 \mpl^2s_{24}}\left(\la \6(k_2-k_4)\5]+\la \5(k_2-k_4)\6]\right)^2$	\\ \bottomrule
    \end{tabular}
    \caption{\footnotesize Amplitudes relevant in the determination of $\Disc F_i$, where $g$ is the gauge coupling (in the normalization of unit charge). Each row corresponds respectively to $X=\phi,\psi,W$.  Other photon helicities are recovered by replacing holomorphic with anti-holomorphic configurations (and vice-versa). Notice, that all amplitudes are given in terms of incoming states, and in order to be used in (\ref{cutkoskirule}) all legs on the r.h.s of the arrows should be flipped by the map $p\rightarrow -p$, and $|\p\ra^{I} \rightarrow -|\p\ra_{I}$, $|\p]^{I} \rightarrow |\p]_{I}$ for massive legs. In this case, the overall effect of the flipping is just the lowering of the $SU(2)$ indices on the massive legs, and no effect on the scalar legs. The contact term proportional to $\xi_\phi$ in the first line is the amplitude counterpart of the scalar non-minimal coupling to gravity, see \eref{nonmincoupX}. Other model dependent contributions, such as those due to Higgs bosons, are discussed in \sref{nonminSec}.  See \aref{conventions} for conventions.}
    \label{amp}
\end{table}

 \begin{table}[t]
    \centering
    \begin{tabular}{c|c|c|ccc}\toprule
$X$ & $\Disc F_1(t\gg m^2)$ & $ \Disc F_2(t\gg m^2)$ &$\Disc F_3(t\gg m^2)$\\	\midrule
$\phi$ & $\frac{i\alpha}{6}$ & ${i\alpha\over 12}(5-4\xi_\phi)\delta(t)$&  $-{i\alpha\over 12}\delta(t)$& \\[0.3cm] 
$\psi$&$ \frac{2i\alpha}{3}$ & $ {i\alpha\over 6}\delta(t)$ &  ${i\alpha\over 6}\delta(t)$ \\[0.3cm] 
$W$& $-\frac{i\alpha}{2}\left( 7- 4\log {t\over m^2}\right)$ &  $-{i3\alpha\over 4}\delta(t)$ &  $-{i\alpha\over 4}\delta(t)$	\\ \bottomrule
    \end{tabular}
    \caption{\footnotesize Limiting behavior of $\Disc F_i$ in the kinematical region $t<0$ and $|t|/m^2\rightarrow \infty$. The Dirac's delta functions signal that the concerned discontinuities vanish pointwise in the massless limit but not under integration.  }
    \label{disclim}
\end{table}

Comparing the tensor structures in \eref{formfactorsAll} or \eref{MatrixTmunu} with the expressions we find for \eref{DiscForm0} using \eref{cutkoskirule} and the amplitudes in \tref{amp}, we extract the form factor discontinuities  $\Disc F_{i}$.   For convenience, we list here $\Disc F_1$ for the three massive spinning particles $\phi,\psi,W$ running in the loop, while the discontinuities of the other form factors are reported in {\aref{FormFactorsExplicit}}  
 \begin{align} \label{AllDiscF1}
& \Disc F_1(t)_{\phi}=  \frac{i \alpha }{6 t^2} \left(t \left(t-10 m^2\right) \sqrt{1-\frac{4 m^2}{t}}+24 m^4 \tanh ^{-1}\sqrt{1-\frac{4 m^2}{t}}\right)\theta(t-4m^2) \\
& \Disc  F_1(t)_\psi=    \frac{2i \alpha }{3t^2} \left(\sqrt{1-\frac{4 m^2}{t}} \left(5 m^2+t\right)t-6 m^2 \left(2 m^2+t\right) \tanh ^{-1}\sqrt{1-\frac{4 m^2}{t}} \right)\theta(t-4m^2) \nonumber  \\ 
& \Disc F_1(t)_W=  \frac{-i \alpha}{2 t^2}  \left( \sqrt{1-\frac{4 m^2}{t}} t\left(10 m^2+7 t\right)-8 \left(m^2+t\right) \left(3 m^2+t\right) \tanh ^{-1}\sqrt{1-\frac{4 m^2}{t}}\right)\theta(t-4m^2) \nonumber 
\end{align}
where $\theta(x)$ is the Heaviside unit-step function and $\alpha=g^2/4\pi$ is the fine structure constant. The $t\gg m^2$ limits of these expressions will be very useful in the following discussion and therefore are listed in \tref{disclim}.

While  $\Disc F_{2,3}(t/m^2\rightarrow \infty)$ vanish pointwise, as expected for the $h=h^\prime=+$ helicities of the photons that forbid any non-trivial products of 4-point amplitudes with exactly massless particles which enter in the unitarity cut, they actually return Dirac $\delta$-functions, see \tref{disclim}. 
A similar effect has been pointed out  in \cite{Korner:1995xd} in the context of the Higgs boson coupling to photons. The connection between these $\delta(t)$ and the IR-side of the trace anomaly is discussed in \sref{traceanomalysect}.

 Notice, moreover, that the constant contribution to $\Disc F_1(t\gg m^2)$ is given by the $\beta$-function of the corresponding particle in the loop, as detailed in \sref{traceanomalysect}. Also, the behavior of the massive spin-1 particle differs from the other contributions by the presence of a $\log^2 t/m$  which will be identified as the contribution of soft divergences in \sref{sudakov}.

\begin{table}[t]\centering
\begin{tabular}{l|lr|cc|cc}
\toprule
\multirow{2}[3]{*}{}  & \multicolumn{2}{|c}{$F_1(t)$} & \multicolumn{2}{|c}{$F_2(t)$}& \multicolumn{2}{|c}{$F_3(t)$} \\
\cmidrule(lr){2-3} \cmidrule(lr){4-5}\cmidrule(lr){6-7}
$X$ & $m^2\gg |t|$ & $m^2\ll |t|$ &$m^2\gg |t|$ & $m^2\ll |t|$ &$m^2\gg |t|$ & $m^2\ll |t|$ \\
\midrule
$\phi$ &  $1+\frac{\alpha  t}{180 \pi  m^2}$ & $1+ \frac{\alpha}{72 \pi }  \left(19-6 \log{-t\over m^2}\right)$ & $ \frac{\alpha (13-10\xi_\phi)}{720 \pi  m^2}$ & $-\frac{\alpha (5-4\xi_\phi)}{24 \pi  t}\!\!$ &  $ -\frac{\alpha }{720 \pi  m^2}$ & $\frac{\alpha }{24 \pi  t}$ \\[0.3cm]
$\psi$  & $ 1+\frac{11\alpha t}{360\pi m^2}$  & $1+\frac{\alpha}{\pi}\left(\frac{35}{36}-\frac{1}{3}\log{-t\over m^2}\right)$ & $\frac{\alpha}{180\pi m^2}$ & $-\frac{\alpha}{12\pi t}$ & $\frac{\alpha}{360\pi m^2}$ & $ -\frac{\alpha}{12\pi t}$ \\[0.3cm]
$W$  & $1+\frac{7 \alpha  t}{20 \pi  m^2}$  & $1- \frac{\alpha}{4\pi}\left( \frac{125}{6} -7 \log{-t\over m^2} +2\log^2{-t\over m^2}\right)$ &  $ -\frac{7 \alpha }{240 \pi  m^2} $ & $ \frac{3 \alpha }{8 \pi  t} $&  $-\frac{\alpha }{240 \pi  m^2}$ & $ \frac{\alpha }{8 \pi  t}$ \\
\bottomrule
\end{tabular}
 \caption{ \footnotesize Large and small $m$ limits of the form factors $F_i$. The EFT parameter $\alpha_3$ in \eref{low_en_lag} is given by   $\alpha_3=-F_3(|t|\ll m^2)$.}
    \label{limFi}
\end{table}

The discontinuities can thus be integrated with the dispersion relations 
\begin{equation}
\label{discAllFs}
F_1(t) =  1+\frac{t}{2\pi i} \int^\infty_{4m^2} \frac{dt^\prime}{t^\prime} \frac{\Disc F_1(t^\prime)}{t^\prime -t-i0^+} \ ,\qquad
 F_{2,3}(t) =  \frac{1}{2\pi i} \int^\infty_{4m^2} dt^\prime \frac{\Disc F_{2,3}(t^\prime)}{t^\prime -t-i0^+}\, ,
\end{equation}
determining  $F_i$ everywhere in the complex cut $t$-plane.  The subtraction constant for $F_1$ has been fixed by the normalization condition \eref{normalizationF1}, so that helicity-preserving low-energy photons scatter gravitationally with strength $1/\mpl$.  The full expressions of $F_i(t)$ are summarized in \aref{FormFactorsExplicit}, while the important limits are collected in \tref{limFi} for convenience. The earliest calculation of $F_{i}$ in spinorial QED was performed in \cite{Berends:1975ah}.

One  particularly interesting limit of \eref{formfactorsAll} is at large masses of the particles running in the loop. This limit is equivalent to integrating out such particles and can be matched to effective irrelevant contributions\footnote{It is actually possible to match as well the form factor contributions in the massless limit but to a non-local 1-loop effective action, using the covariant effective action approach of e.g. \cite{Barvinsky:1985an,Barvinsky:1994hw}.}
\bea\label{low_en_lag}
\mathcal{L}=-{1\over 4}F_{\mu\nu}F^{\mu\nu}+\alpha_1 R F_{\mu\nu}F^{\mu\nu}+\alpha_2 R_{\mu\nu}F^{\mu\alpha}F^{\nu}_{\alpha}+\alpha_3 R_{\mu\nu\alpha\beta}F^{\mu\nu}F^{\alpha\beta}+ \dots  \, ,
\eea
of which we only display the off-shell 3-point vertices.  Notice that only $\alpha_3$ contributes to the on-shell 3-point function $\gamma\gamma-\mathrm{graviton}$, and the $\alpha_{1,2}$ correct instead  low-energy on-shell 4-point amplitudes only, as is visible by using the equations of motion.   We remark that the form factor $F_3$ reduces to the Wilson coefficient of $F_{\mu\nu}F_{\alpha\beta}R^{\mu\nu\alpha\beta}$ i.e.  $\alpha_3=-F_3(t\ll m^2)$, which give rise to  on-shell  helicity-violating 3-point vertex $\gamma\gamma-\mathrm{graviton}$, at low energy.   The  electron and scalar Wilson coefficients for $\alpha_3$ nicely agree with the results present in the literature, see e.g. \cite{Berends:1975ah,Drummond:1979pp,Cheung:2014ega} and references therein.   
The effective Wilson contribution to $\alpha_3$ from massive vectors is new to the best of our knowledge.

\subsection{Non-Minimal Couplings and Higgs/Graviton Mixing}
\label{nonminSec}

The ($\xi_\phi$-independent part of the) form factors we have calculated are generated at one loop by charged states minimally coupled to photons and gravity.
Within this setup neutral particles contribute from 2-loop order only. 
With non-minimal couplings, instead, other 1-loop contribution are generated, even from neutral scalars. 

In this subsection we discuss two illustrative cases of non-minimal gravitational coupling for charged $(\phi)$ and neutral ($H$) scalars. 
The latter is actually relevant because of the Higgs {mechanism} whenever charged spin-1 particles are considered, should the Higgs boson be coupled to gravity non-minimally.  
We report the result for the Standard Model Higgs at the end of this subsection. 

Let us consider  non-minimal gravitational couplings  given by 
\begin{equation}
\label{Snonmin}
S \supset \int d^4 x \sqrt{|g|} \frac{R}{6} \left(\xi _\phi |\phi|^2 + \frac{\xi_H}{2} H^2\right)  \Longrightarrow \delta T_{\mu\nu}=-\frac{1}{3}\left(\partial_\mu\partial_\nu-\eta_{\mu\nu}\square \right)\left(\xi_\phi |\phi|^2+ \frac{\xi_H}{2}H^2\right)\,,
\end{equation}
where we extracted the energy-momentum tensor from the linear gravitational coupling around Minkowski, that is $\delta S=\int d^4 x \sqrt{|g|} T_{\mu\nu} \delta g^{\mu\nu}/2$.  They clearly contribute to identically conserved improvement terms of the energy-momentum tensor,  hence changing the  $F_2$ form factor  in  \eref{formfactorsAll}. The case with $\xi_{\phi} = \xi_H = 1$ and vanishing masses is known as conformally coupled scalars because the two-derivative action becomes classically Weyl invariant.

One simple way to take the effect of $\xi_{\phi,H}$ into account, which makes also direct contact with the on-shell method approach we have taken in the rest of this work,  is by removing the non-minimal couplings via a field redefinition that is effectively equivalent to plugging the unperturbed equation of motion $R=-T_\mu^\mu/\mpl^2$ in the action in \eref{Snonmin}.\footnote{See \aref{HiggsGravitymix} for an equivalent discussion phrased in terms of Higgs/graviton mixing resolved by field redefinitions. }  This gives rise to new contact-term interactions associated to the trace of the energy-momentum tensor 
\begin{equation}
\label{redefSnonmin}
S \supset  \int d^4 x  \sqrt{|g|} \frac{1}{6\mpl^2}\left(\xi _\phi |\phi|^2 + \frac{\xi_H}{2} H^2\right)\left[(\partial S)^2+2|\partial\phi|^2+(\partial H)^2+\ldots \right]\,.
\end{equation}
Therefore, the effect of non-minimal coupling associated to $\xi_\phi$ is nothing but changing the on-shell data that enter in the calculation of the discontinuities, i.e. 
\begin{equation}
\label{nonmincoupX}
\delta \cM(5_\phi 6_{\bar \phi}\rightarrow 2_S 4_S)=  -\frac{\xi_\phi}{6} \frac{s_{24}}{\mpl^2}  \ , 
\end{equation}
as reported in the first row of \tref{amp},  the first row of \tref{limFi}, and more generally in the $F_2$ reported in \aref{FormFactorsExplicit}. 

On the other hand, $\xi_H$ does not affect $F_2$ at one loop (to $O(1/\mpl^2)$) unless $H$ gets a VEV $\langle H \rangle=v$  and it couples to charged states running in the loop, both conditions being actually satisfied for the neutral component for the Higgs boson field of the Standard Model. 
Indeed, from the $\delta T_{\mu\nu}$ in \eref{Snonmin}, after replacing the perturbations around the VEV, we have 
\begin{equation}
\label{Higgscontribution2}
\langle 0| \delta T^{(H)}_{\mu\nu}(0) |\gamma\gamma\rangle=\frac{\xi_H v}{3}\left(q_\mu q_\nu-\eta_{\mu\nu} q^2 \right)\frac{-1}{s_{13}-m_H^2}\cM(1_\gamma 3_\gamma \rightarrow H)
\end{equation}
where $m_H$ is the Higgs boson mass, and the amplitude $\cM(1_\gamma 3_\gamma \rightarrow H)$ is model dependent. The latter depends on the  trilinear Higgs boson coupling $HX\bar{X}$ where $X$ is  running in the loop.  \eref{Higgscontribution2} produces as well a shift in the on-shell scattering data between the Higgs and the spectator field 
\begin{equation}
\label{Higgscontribution}
\delta\cM(1_\gamma 3_\gamma \rightarrow 2_S 4_S)=\cM(1_\gamma 3_\gamma \rightarrow H) \left(\frac{s_{13}}{s_{13}-m_H^2}\right) \left(\frac{v\xi_H}{6\mpl^2} \right)
\end{equation}
as one can also check directly from \eref{redefSnonmin}.

Let's put these expressions to good use and consider the example of $SU(2)\rightarrow U(1)$ symmetry breaking pattern for a  weakly coupled $SU(2)$ gauge theory where a real triplet $\vec\phi$ gets the VEV $\langle \phi^{i} \rangle=\delta^{i3} v$. The low-energy physical spectrum contains a massless photon and, since the theory is weakly coupled, it contains as well  a pair of spin-1 bosons $W^{\pm}$ of mass $m_W^2=(gv)^2$  and a neutral Higgs boson $H$ in $\vec{\phi}=(\pi^1,\pi^2,v+H)$.  For convenience, we now call $H$ the fluctuation around the VEV.  Everything we discussed in this subsection applies directly to $\phi_3$  in the unitary gauge where $\int d^4 x \xi_HR\vec{\phi}^2/12 = \int d^4 x \xi_HR (v+H)^2/12$.  The amplitude  in the $s_{13}\gg m_W^2$ and $s_{13}\ll m_W^2$ limits is extracted immediately via the Goldstone equivalence theorem (see e.g. \cite{Chanowitz:1985hj,Horejsi:1995jj}) and the Higgs low-energy theorem \cite{Shifman:1979eb}, respectively, 
\begin{align}
\begin{split}
& \frac{\cM(1_\gamma 3_\gamma \rightarrow H)}{ \left[2(\epsilon^\prime \cdot k)(\epsilon \cdot k^\prime)- q^2 (\epsilon \cdot \epsilon^\prime)\right]}\big|_{s_{13}\gg m_W^2}=  -\frac{2}{v} \left(\frac{\alpha}{4\pi }\right) \,, \\  
& \frac{\cM(1_\gamma 3_\gamma \rightarrow H)}{\left[2(\epsilon^\prime  \cdot k)(\epsilon \cdot k^\prime)- q^2 (\epsilon \cdot \epsilon^\prime)\right]}\big|_{s_{13}\ll m_W^2}= -\frac{7}{v} \left(\frac{\alpha}{4\pi }\right) 
\end{split}
\end{align}
so that \eref{Higgscontribution2} compared to \eref{formfactorsAll} returns 
\begin{align}
\label{deltaF2limitHiggs}
\delta F_2(t)\big|_{t\gg m_W^2} =   \frac{2\xi_H}{3} \left(\frac{\alpha}{4\pi }\right) \frac{1}{t-m_H^2}\,,\qquad \delta F_2(t)\big|_{t\ll m_W^2} =   \frac{7\xi_H}{3} \left(\frac{\alpha}{4\pi }\right) \frac{1}{t-m_H^2} \ ,
\end{align}
where we remind the reader that $s_{13}=t$.  The $t\gg m_{H,W}^2$ limit of $t \, \delta F_2(t)$ in \eref{deltaF2limitHiggs} enters directly in the trace anomaly equation we  will discuss in  \sref{traceanomalysect}.  

Next we move to the result for $\delta F_2$ valid for all $s_{13}$.  Since this model contains the same spectrum and couplings of the particles that generate the spin-1 contribution to the  {$H \rightarrow \gamma\gamma$} process in the Standard Model, we can directly use  the $W$-boson contribution from the Standard Model expression of $\cM(1_\gamma 3_\gamma \rightarrow H)$ which, incidentally can also be extracted by dispersion relations and on-shell data  \cite{Melnikov:2016nvo} being careful with the subtraction constants that can be fixed by matching to the Goldstone equivalence theorem and the Higgs low-energy theorem results.  From the Standard Model $W$-contribution to $\cM(1_\gamma 3_\gamma \rightarrow H)$  we thus get 
\begin{equation}
\label{F2frommixing}
\delta F_2 (t) = \frac{2\xi_H }{3} \frac{\left(\frac{\alpha}{4\pi}\right) }{t-m_H^2} \left\{   1+6\frac{m_W^2}{t} -3\frac{m_W^2}{t}\left(1-2\frac{m_W^2}{t}\right)  \left( \log \frac{1+\sqrt{1-4m_W^2/t}}{1-\sqrt{1-4m_W^2/t}} -i\pi \right)^2 \right\}\,.
\end{equation}
A similar contribution from the top-quark  can easily be included as well in $\cM(1_\gamma 3_\gamma \rightarrow H)$, so that \eref{Higgscontribution2} can be used to determine $\delta F_2$ contribution from the Higgs boson of the Standard Model.

\subsection{Trace Anomaly and the Running Coupling}
\label{traceanomalysect}
The trace of the energy-momentum tensor is 
\begin{align}
\begin{split} \label{traceF2F3}
\langle 0| T^\mu_\mu(0)  |k^{\prime h^\prime} k^{h}\rangle= & -t\left( 3F_2(t)+F_3(t)\right)  \left[2(\epsilon^\prime \cdot  k)(\epsilon \cdot k^\prime)- q^2 (\epsilon \cdot \epsilon^\prime)\right] \\ 
= & \, t \left( 3F_2(t)+ F_3(t)\right) 
\left(
\begin{array}{cc}
0    & \langle k k^\prime \rangle^2 \\ 
 \, \!  [k k^\prime]^2 & 0 \\
\end{array}
\right) \ ,
\end{split}
\end{align}
which depends only on the the combination $t(3F_2+F_3)$ and is non-zero only for helicity-flipping photons, as it should be for the overlap of photons with the spin-0 state $ \langle 0 | T_{\mu}^\mu(0) $. 

For $t\gg 4m^2$ and $t\gg m_H^2$, i.e. well above the pair production threshold of massive charged particles and Higgs bosons, and setting $\xi_{\phi} = \xi_H  =1$ for all conformal couplings in \eref{redefSnonmin} such that the dilation current can be written as $ x_\nu T^{\mu \nu}$, the QED trace anomaly 
\begin{equation}
T^\mu_\mu= \frac{\beta}{2 g} F_{\rho\sigma}^2\,, 
\end{equation}
 implies $\langle 0| T^\mu_\mu  |k^\prime k\rangle \mathcal{N}=\frac{\beta}{g} \left[2(\epsilon^\prime \cdot  k)(\epsilon \cdot k^\prime)- q^2 (\epsilon \cdot \epsilon^\prime)\right]$. Therefore \eref{traceF2F3} allows us to express the $\beta$-function for the gauge coupling $g=g(\mu)$ in terms of the form factors
\begin{equation}
\label{betafunF2F3}
\beta=-g \lim_{t/m^2\rightarrow \infty} t\left[3F_2(t)+F_3(t)\right]_{\xi_{\phi,H} =1}\,.
\end{equation}
From the explicit expressions that we calculated, see \tref{limFi} and \eref{deltaF2limitHiggs}, we can read off the QED $\beta$-functions from loops of charged spin-0, spin-1/2 and spin-1 particles (for unit charges) as
\begin{equation}\label{betai}
\beta_{\phi}=\frac{1}{3} \left(\frac{g^3}{16\pi^2}\right) \,,\qquad \beta_{\psi}=\frac{4}{3} \left(\frac{g^3}{16\pi^2}\right) \,,\qquad \beta_{W}=-7\left(\frac{g^3}{16\pi^2}\right) 
\end{equation} 
In $\beta_{W}$ we included the contribution from the Higgs/graviton mixing, \eref{deltaF2limitHiggs}. 

It is a highly non-trivial result that from gravitational form factors we correctly reproduce the non-abelian (negative) $SU(2)$ $\beta$-function $\beta_{W}/(g^3/16\pi^2)= -11/3 \times 2 + 1/6 \times 2 = -7$, including the scalar matter in the adjoint representation, and using purely on-shell  data associated to scattering only physical polarizations. This connection between the energy-momentum tensor, scattering data, and the $\beta$-functions is somewhat reminiscent of the methods presented in \cite{Caron-Huot:2016cwu}.

We remark that the finite value at $t/m^2\rightarrow \infty$ of the $\beta$-functions as calculated by the trace anomaly boils down to the presence of an IR-localized Dirac $\delta$-function in $\Disc F_{2,3}$ that we have reported in \tref{disclim}. These Dirac $\delta$-functions represent the IR side of the trace anomaly in full analogy with the chiral anomaly case, see \cite{Dolgov:1971ri,Coleman:1982yg}, as it was already pointed out for spinorial QED in \cite{Giannotti:2008cv}.  
By contrast,  we explore below the UV side associated to the running coupling and expose its connection to the $F_1$ form factor. 

By putting the theory on a  curved spacetime\footnote{The gauge coupling $g=g(\mu)$ should not be confused with the metric determinant in the volume element $\sqrt{|g|}d^4x$.}  
 \begin{equation}
\label{QEDcurved}
S = \int d^4x \sqrt{|g|}  \left\{ -\frac{1}{4g^2(\mu)}F_{\mu\nu}F^{\mu\nu} -\frac{\mpl^2}{2} R+\ldots \right\} 
\end{equation}
we can as well establish an important connection between the $\beta$-function and the $F_1$ form factor which is directly connected to the asymptotic time delay at short impact parameter, as we show in \sref{Phaseshiftsection}.  Expanding the action \eref{QEDcurved} to first order in the metric perturbations around Minkowski spacetime and Fourier-transforming the photon field (with a slight abuse of notation) $\epsilon_\mu(k)=\int d^4x e^{ikx} A_\mu(x)$ we get  
\begin{align}
 S \supset    - \int \frac{d^4q}{(2\pi)^4}  \frac{d^4k^\prime}{(2\pi)^4} \frac{d^4k}{(2\pi)^4} (2\pi)^4\delta^4(q+k+k^\prime) h^{\mu\nu}(q) \frac{1}{2g^2(\mu)} \langle 0 | T_{\mu\nu}(0) | k^\prime k \rangle\big|^{\mathrm{tree}}\mathcal{N}+\ldots  
   \end{align}
The same running coupling $g=g(\mu)$ in front of the photon kinetic term is found as well in the $F_1$ form factor. That is, the counter-term needed to renormalize the photon kinetic term also enters in the $F_1$ form factor. Therefore, in the limit $q^2\gg m^2$,  rather then defining $g(\mu)$ as the coupling of  an off-shell $A_\mu$ to charged currents, as e.g. measured in the Coulomb potential at a floating renormalization scale $-q^2=\mu^2$  (i.e. in a scattering process mediated by a virtual photon),  we can equally think of it as the coupling of two incoming helicity-preserving on-shell photons scattering on an off-shell graviton with $-t=\mu^2$
\begin{equation}
\label{F1beta}
\frac{d}{d\log\mu} \frac{1}{g^2(\mu)}= \frac{d}{d\log\mu} F_1(t=-\mu^2)\big|_{m\ll\mu\,, g=1} \Longrightarrow \beta=-\frac{g}{2} \frac{d}{d\log\mu} F_1(t=-\mu^2)\big|_{m\ll\mu} \,. \end{equation}
In the right-most expression in \eref{F1beta}  we have restored to a canonically normalized kinetic term $-1/4 F_{\mu\nu}^2$ in the lagrangian density.  
The formula in \eref{F1beta} links directly the sign of the $\log(-t)$ in the helicity-preserving form factor $F_1$ to the sign of the $\beta$-function.\footnote{Trading the $\log\mu^2$ dependence for the $\log q^2$  breaks down, however,  if extra $\log q^2/m^2$ factors survive in the $q^2 \gg m^2$ limit, which signals the presence of IR divergences. They do not arise at one loop of spin-0 and spin-1/2 charged states, but are instead present for spin-1 particles for which, therefore, \eref{F1beta} and \eref{betalimitF1} no longer apply. We study IR divergences in \sref{sudakov}.  }
 Moreover, from the dispersion relations \eref{discAllFs} the $\log ( - t )$ arises, in the case of spinorial and scalar QED, by the constant limit of the discontinuity $\Disc F_1(t\rightarrow\infty)$, hence 
\begin{equation}
\label{betalimitF1}
\beta_{\phi,\,\psi}=\lim_{t/m^2\rightarrow \infty}\frac{g}{\pi}\frac{\Disc F_{1}(t)_{\phi , \psi}   }{2i}\,.
\end{equation}
This nice expression connects directly the $\beta$-function in spinorial and scalar QED  to the discontinuity of the helicity-preserving gravitational form factor $F_1$,  i.e. to on-shell-only gravitational scattering amplitudes. From the first two rows of the first column of \tref{disclim} one indeed reproduces the $\beta_\phi$ and $\beta_{\psi}$ in \eref{betai}.

\subsection{IR-Divergences and Sudakov Double-Logarithms}\label{sudakov}

The presence of a finite ---but still large--- $\log^2$ factor in the high energy limit of the helicity-preserving form factor $F_1$ generated  at one loop by massive charged spin-1 $W$ bosons 
\begin{equation}
\label{highlimitF1J1}
 \frac{\langle 0| T_{\mu\nu}(0)|1_\gamma^- 3_\gamma^+\rangle^{(1\text{-loop})}}{\langle 0| T_{\mu\nu}(0)| 1_\gamma^- 3_\gamma^+ \rangle^{\mathrm{tree}}} \xrightarrow[t/m_W^2 \rightarrow \infty]{}  1- \frac{\alpha}{4\pi}\left( \frac{125}{6} -7 \log{-t\over m_W^2} +2\log^2{-t\over m_W^2}\right)\,
\end{equation}
can be understood as the IR divergences that we would encounter  if the $W$ mass were vanishing. 
It arises  in a way that is  completely analogous to the presence of large double-logarithms in the matrix elements of electroweak currents, see e.g. \cite{Ciafaloni:1998xg,Fadin:1999bq,Ciafaloni:2009tf}, which are usually called electroweak Sudakov double-logarithms  in analogy to the original QED Sudakov factors that are associated to the vanishing photon mass. 

We deal with these Sudakov factors in the self-energy by taking a renormalization group approach to resum the leading double-$\log$ factors. 
We first regulate the most IR-singular class of diagrams by cutting them with a floating mass $m_W\rightarrow m=\mu$, which should be taken not too far from the kinematical variables  so that perturbation theory is reliable,  and then we evolve the form factor down with the resulting RG equation that we can thus write as an evolution in the mass, namely  
\begin{equation}
\label{evolutionEqSudakov}
\frac{\partial \langle 0| T_{\mu\nu}(0)| 1_\gamma^- 3_\gamma^+ \rangle }{\partial\log (m^2/(-t))}= -4\left(\frac{\alpha}{4\pi}\right) \log\left(\frac{m^2}{-t}\right)\, \langle 0| T_{\mu\nu}(0)| 1_\gamma^- 3_\gamma^+ \rangle\,. 
\end{equation}
Integrating this RG equation from $\mu=\mu_0$ down to the $W$ mass $\mu = m_W$ we get the exponentiation of the Sudakov double-$\log$s in the form factor 
\begin{equation}
F_1(t) \simeq F_1(\mu_0^2) \mathrm{Exp} \left[-2\left( \frac{\alpha}{4\pi}\right) \left(\log^2 \frac{m_W^2}{-t}-\log^2 \frac{\mu_0^2}{-t}\right)\right]
\end{equation}
and the associated exponential suppression for $t\gg m_W^2$ 
\begin{equation}\label{sudakF1larget}
|F_1(t\gg m_W^2)| \propto \left(|t|/m_W^2\right)^{-2(\alpha/4\pi )\log|t|/m_W^2} \to 0
\end{equation}
 at high-energy. 
 
We remark that, while we have obtained the evolution equation \eqref{evolutionEqSudakov} within perturbation theory, it holds in fact non-perturbatively as shown in \cite{Fadin:1999bq}. The exponential  suppression from the leading double logs is indeed completely fixed by the sum of the quadratic Casimirs $S_i(S_i+1)$ associated to the representations, carried by each $i$-th external leg, of the $SU(2)$ gauge group (as we are dealing with a non-abelian gauge theory with just $W^{\pm}$ and $\gamma=W^3$ in the spectrum).  The evolution equation of \cite{Fadin:1999bq} is indeed nicely matched by our perturbative derivation, \eref{evolutionEqSudakov}. 

Notice that the same exponentiation of the IR Sudakov logs  takes place for the helicity-flipping matrix elements $F_{2,3}$, but starting at two-loop order $O(\alpha^2)$. It can  be obtained by adapting again the results of e.g. \cite{Fadin:1999bq}, something that we leave to future investigations, limiting the present work to 1-loop accuracy.
 
While the exponential suppression we find is analogous to the vanishing of exclusive processes in ordinary QED,  here  the $W$ mass is finite and this makes the resummed form factor and the associated exclusive amplitudes actually finite. Moreover, the finiteness of the mass and charge of the $W$ boson allows one to distinguish states with different numbers of $W$ particles in them, contrary to the case of photon emissions which can always escape detections if sufficiently soft. For these reasons, we keep working in the following with the exclusive 2-photon matrix elements of $T_{\mu\nu}$, which is thus IR finite since charged particles are consistently excluded in the final or initial state, rather than with inclusive cross-sections.  
Moreover, the effect of the exponential suppression  is relevant only for $(\alpha/2\pi ) \log^2\left( -t/m_W^2\right) \gg 1$, and it is therefore not important to the phase shift \eref{deltacombined} in the region $1/m_W^2 \mathrm{exp}(-\sqrt{2\pi/\alpha})\ll b^2 \ll 1/m_W^2$, i.e. for impact parameters that can still be taken exponentially smaller than $1/m_W^2$ at weak coupling. In the following sections we consistently include the impact of the exponentiation in the regime $b^2 m_W^2 \ll \mathrm{exp}(-\sqrt{2\pi/\alpha})$. 

It would nevertheless be interesting to study in a future work the inclusive case where the double-log contributions would cancel out so to become sensitive again to the single logs and possibly to  the sign of the beta-function, like it is the case for spinorial and scalar QED when not embedded in Yang-Mills theory. 

\section{{The Phase Shift}}
\label{Phaseshiftsection}
Now that the form factors $F_i(t)$ have been determined, we proceed to computing various quantities of interest. We study the 4-point function (diagrammatically shown in \fref{Fi}) in the eikonal limit $s\gg t$, where the center of mass energy is much larger than the exchanged momentum, corresponding to the  response of photons to the gravitational field generated by the spectator fields. 

In this limit, and scattering either at transplanckian center of mass energy or against several spectators, the amplitude exponentiates in impact parameter space,  $S=e^{2i \delta(s,b)}$ where $\delta(s,b)$ is referred to as the phase shift and $b$ is the impact parameter. The large phase in the eikonal transplanckian scattering against a single spectator is generated when $G s\gg 1$ and $G\sqrt{s}\ll b$, see e.g. \cite{tHooft:1987vrq,Amati:1987uf,Amati:1992zb,Kabat:1992tb}, while with several $N$ shock-waves  each scattering is subplanckian building up to $Ns G\gg1$  \cite{Camanho:2014apa}.  
 The phase shift is related to a number of observables such as scattering angle and the time delay, as explicitly shown in \sref{timedelay}.  
 In this section we present the leading quantum corrections in the gauge coupling to the phase shift in the eikonal regime. 
 
 The 1-loop quantum corrections from pure gravity, $\delta_1\sim G^2 s/b^2$ \cite{Amati:1987uf},  are always very small in the transplankian eikonal regime, set in fact  by the ratio of the Planck length over the impact parameter $b$.  They are also much smaller than the 2-loop gravitational corrections $\mathrm{Re}\delta_2\sim G s(R_s/b)^2$. The 1-loop gauge contribution scales instead as $G s (\alpha/4\pi) \log^2(m_X b)$ for $b<1/m_X$ and $Gs  (\alpha/4\pi) / (  m_X b )^2$ for $b>1/m_X$, see \eref{deltasmallblimit} and \eref{deltalargeblimit2},  which can be much larger than 1 and dominate over the gravitational $\delta_2$ for a suitable range of $s$ and $b$ we restrict to. Similar scaling applies to the case of scattering against a coherent spectator background.

\subsection{Amplitudes in the Eikonal Limit}
In this section, we present the eikonal limit of the 4-point amplitude, which will be used in the evaluation of the time delay, following \cite{Camanho:2014apa,AccettulliHuber:2020oou}. For simplicity, we detail the construction for a scalar spectator, but we have checked that in the eikonal limit, the same result is obtained by scattering against spin-1 and spin-2 spectators minimally coupled to gravity. In other words, the spin of the spectator is irrelevant in the computation of the time delay, as long as  it is characterized only by a minimal gravitational interaction. 

 When contracted with the scalar 3-point function, the full amplitude takes the form 
\bea\label{full_scal_mat}
\cM (1_\gamma 3_\gamma \rightarrow 2_S 4_S) = \begin{pmatrix}
 -{\la 3  k_2 1]^2\over\mpl^2 s_{13}}F_1(t)  & {\la 13 \ra^2\over 2\mpl^2}\left(  s_{13} F_2(t)+{(s_{12}-s_{14})^2\over s_{13}}F_3(t) \right) \\
{[ 13 ]^2\over 2\mpl^2}\left( s_{13} F_2(t)+{(s_{12}-s_{14})^2\over s_{13}}F_3(t) \right) &   -{\la 1 k_2 3]^2\over\mpl^2 s_{13}}F_1(t)   
\end{pmatrix} \ , 
\eea
where we recall that the diagonal entries correspond to the helicity preserving amplitudes with $h^\prime=-h=\pm $  while the off-diagonal entries correspond to the helicity-flipping $h=h^\prime=\pm $,  and the Mandelstam variables $s_{ij}$ are defined in \aref{conventions}.
 This amplitude is evaluated on the following massless kinematic configuration
\begin{align}
\begin{split} \label{kin_Mal}
& k_1^\mu=(\omega,-\vec{p}+{\vec q / 2})\ ,\qquad k_3^\mu=-(\omega,-\vec{p}-{\vec q / 2})\ ,\\
& k_2^\mu=(\omega,\vec{p}-{\vec q / 2})\ ,\qquad k_4^\mu=-(\omega,\vec{p}+{\vec q / 2})\ ,
\end{split}
\end{align}
where $\vec q$ is the exchanged momenta, $\omega =\sqrt{\vec{p}^{\,2}+\vec{q}^{\,2}/4}$, and in the following we fix the direction of $\vec p=p \hat{z}$, where $\hat{z}$ is the unit vector in the $z$-direction. The Mandelstam variables in this configuration are given by 
\bea
s=s_{12}=4\omega^2, \quad t=s_{13}=-\vec q^{\,2}, \quad u=s_{14}=-4 \vec p^{\,2} \ . 
 \eea
By momentum conservation, the product $\vec p \cdot \vec q$ is zero, implying that the momentum transfer $\vec q$ lies in the $xy$-plane. With an abuse of notation, we will refer to $\vec q$ as a two-dimensional vector with components $\vec q = (q_1,q_2)$.

We are interested in the eikonal approximation $\omega \gg |\vec q\,|$, where the amplitude \eref{full_scal_mat} in the kinematic configuration \eref{kin_Mal} is given by
\bea\label{eik_amp}
\cM^{\mathrm{eik}}(t)=\frac{s^2}{\mpl^2 \vec{q}^{\,2}}
\begin{pmatrix}
F_1(t) & -4 q_+^2F_3(t)  \\[0.3cm]
-4q_-^2 F_3(t)& F_1(t) 
\end{pmatrix}\ ,
\eea
where $q_+={1\over\sqrt{2}}(q_1+iq_2)$ and $q_-={1\over\sqrt{2}}(q_1-i q_2)$, and we dropped the contribution from $F_2(t)$ which is analytic in $t$, hence giving rise, once Fourier transformed to impact parameter space $b$, only to local terms such as $\delta(b)$ or derivatives thereof.  This means that the improvement {terms proportional to $\xi_{\phi}$ and $\xi_H$} do not produce any measurable effect on the time delay. 

\subsection{Computation of the Phase Shift}
\label{compPhaseShift}
The phase shift is obtained by Fourier transforming the 4-point amplitude in the eikonal limit \eref{eik_amp} to impact parameter space  
\bea
\delta(s,b)={1\over 4 s } \int \frac{ d^2 q }{(2 \pi)^2}  e^{i\vec b\cdot \vec q}\cM^{\mathrm{eik}}(t=-\vec{q}^{\,2})\ ,
\eea
where $b \equiv | \vec{b}|$.  
The eigenvalues of this matrix are given by
\bea
\label{eigenvaluesPhaseShift}
\delta_\pm(s,b )={s\over 4 \mpl^2}\left[ \hat F_1(b^{2}) \pm16 b^{2} \hat F_3''(b^{2}) \right]\ ,
\eea 
where we have defined
\bea
\label{Fihat}
\hat F_i   (b^{2})  =\int \frac{d^2 q}{(2 \pi)^2} {F_i(-\vec{q}^{\,2})\over \vec{q}^{\,2}}e^{i\vec{b}\cdot \vec{q}} \ ,
\eea
for $i = 1,3$, and {$\hat F_3^\prime \equiv \partial \hat F_3 / \partial b^{2}$}.

\begin{figure}[t]
\centering
\includegraphics[width=8cm]{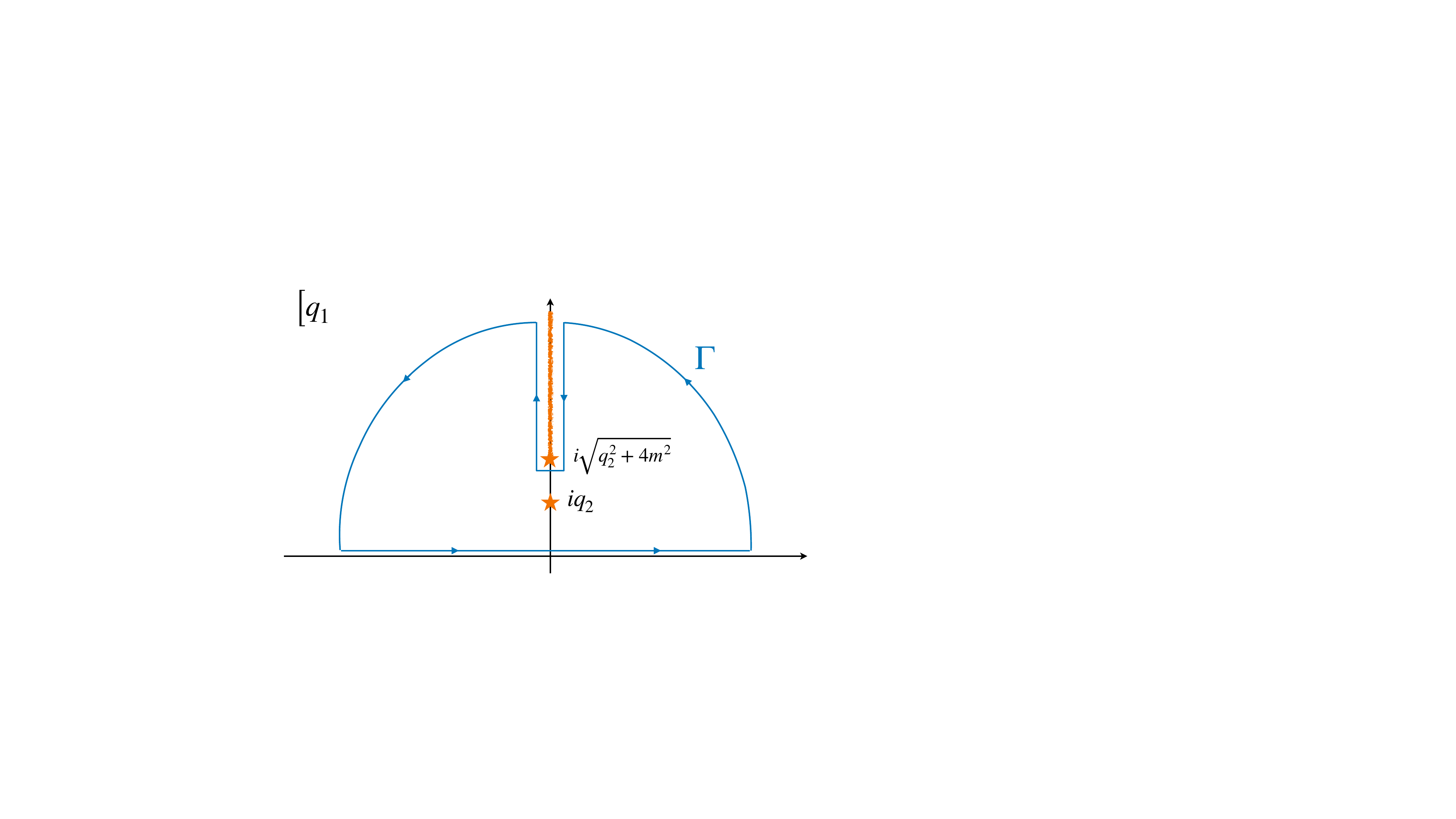}
\caption{ \footnotesize Integral contour $\Gamma$ in the upper complex $q_1$-plane for $\hat F_i$. There are two contributions: one from the graviton pole, and the second from the discontinuity above threshold $t>4 m^2$. }
\label{ana_Fi}
\end{figure}

The integrand in Eq.~\eqref{Fihat} is discontinuous at the graviton pole or above threshold, i.e. when $q_1 = \pm i q_2$ or $t=-\vec{q}^{\,2} > 4m^2$. We can then compute Eq.~\eqref{Fihat} by applying the Cauchy theorem. The integration contour of $q_1$ can be deformed in its complex plane so to express the integral in terms of the discontinuities computed in \tref{disclim}, see \fref{ana_Fi}. Without loss of generality, we can fix $\vec{b}=(b,0)$ because of rotation invariance. After performing the rotation $q_1 = i Q_1$ and changing the order of integration, Eq.~\eqref{Fihat} takes the form
\begin{equation}\label{poledisc}
\hat F_i(b^{2})= -{F_i(0)\over 2\pi}\log{b/b_{\mathrm{IR}}} +\frac{2i}{(2\pi)^2}\int_{2m}^{+\infty}d Q_1 \int_0^{\sqrt{Q_1^2-4m^2}}dq_2\frac{\Disc{F}_i(Q_1^2-q_2^2)}{Q_1^2-q_2^2}e^{-Q_1 b}\ ,
\end{equation}
where $b_{\mathrm{IR}}$ is an infrared cutoff. The $b_{\mathrm{IR}}$ has no physical impact as long as one considers wave-packets with $b<b_{\mathrm{IR}}$.     The integral in \eref{poledisc} can be further simplified by changing variables $Q_1 = \sqrt{t}\cosh\theta,q_2 = \sqrt{t}\sinh\theta$, in terms of which it becomes
\begin{equation}\label{polediscv3}
\hat F_i(b^{2})= -{F_i(0)\over 2\pi}\log{b/b_{\mathrm{IR}}} + \frac{i}{(2\pi)^2}\int_{4m^2}^{+ \infty} dt  \frac{\Disc{F}_i(t)}{t}  K_0\left(b\sqrt{t}\right) \,,
\end{equation}
where $K_0 $ is the modified Bessel function of the second kind.  Combining \eref{polediscv3} with \eref{eigenvaluesPhaseShift}, we obtain the final expression for the phase shift
\begin{align}
\begin{split} \label{deltacombined}
\delta_{\pm} ( s , b ) & = \frac{s}{4 \mpl^2} \Bigg[  - \frac{1}{2 \pi} \left( F_1 ( 0 ) \log \frac{b}{b_{\rm IR}} \mp \frac{8}{b^2} F_3 ( 0 ) \right) \\
& \hspace{.8in} + \frac{i}{(2\pi)^2} \int_{4 m^2}^{+\infty} \frac{dt}{t} \left( \Disc F_1(t) K_0 \left(b \sqrt{t} \right) \pm 4 \, t \, \Disc F_3 ( t ) K_2 \left( b \sqrt{t} \right)    \right)  \Bigg] \  ,
\end{split}
\end{align}
which makes manifest that the phase shift $\delta(s,b)$ depends just on the {$t \rightarrow 0 $ graviton pole and the} $t$-channel discontinuities of the self-energy form factors, i.e. on-shell data.

Before discussing the whole 1-loop calculation, we focus on the tree-level contribution, which corresponds to $F_1(t)=1$ and $F_2(t)=F_3(t)=0$. In this case, \eref{eigenvaluesPhaseShift} and \eref{polediscv3} return the tree-level contribution to the phase shift as
\bea \label{treelevelphase1}
\delta_{0}(s,b)=-{s\over 8\pi \mpl^2}\log{b/b_{\mathrm{IR}}}\,.
\eea
Since the IR cutoff $b_{\mathrm{IR}}$ is the largest length scale that we consider,  \eref{treelevelphase1} always leads to a positive contribution to the phase shift.  At 1-loop, there are additional contributions coming from $F_i(0)$ and the discontinuity, see \tref{limFi}. In the following two sections, we study \eref{polediscv3} analytically in two opposite regimes in parameter space:  $b\gg 1/m$ and $b\ll 1/m$, while the full solution is solved numerically and displayed in \fref{plots}.

\subsubsection{The Large $b$ Limit}
In the scenario $b\gg 1/m$, we can use the asymptotic behavior of the Bessel function $K_0(b\sqrt{t}) \sim e^{-b \sqrt{t}}/\sqrt{b t^{1/2}}$ which shows that the contribution from the integral over the discontinuity is exponentially suppressed. Therefore, the only contribution comes from the graviton pole, and the phase shift is given by
\bea
\label{deltalargeblimit2}
\delta_\pm(s , b\gg 1/m)=\delta_{0}(s,b) \pm {s F_3(0) \over \pi \mpl^2 b^2}\ ,
\eea
where $F_3(0)$ is summarized in \tref{limFi} for different spins of the particle in the loop. This is the result one would obtain by working in the EFT where the massive states have been integrated out, and it reproduces the correction from the effective term $F_{\mu\nu}F_{\alpha\beta}R^{\mu\nu\alpha\beta}$ computed first in \cite{Camanho:2014apa} and discussed at the end of \sref{oneloopdisc} around \eref{low_en_lag}. Notice that the only contribution from $F_1(t)$ comes from the tree-level amplitude, as all corrections vanish when evaluated on the pole.

\subsubsection{The Small $b$ Limit for Scalar- and Fermion-loops}
We can write the integral in \eref{polediscv3} in terms of a dimensionless variable $y=b\sqrt{t}$
\begin{equation}
\hat F_i(b^2)= -{F_i(0)\over 2\pi}\log{b/b_{\mathrm{IR}}} +\frac{i}{2\pi^2}\int_{2mb}^{+\infty} \frac{dy}{y}  \Disc{F}_i(y^2/b^2)\,K_0\left(y\right) \,.
\end{equation}
Let us first focus on the helicity preserving contribution $F_1(t)$ to the phase shift in \eref{eigenvaluesPhaseShift}.
In the small $b m$ regime, the integrand $\Disc F_1(y^2/b^2)$, which is actually a function of the dimensionless ratio $y^2/b^2m^2$, receives contribution mostly from the $\Disc F_i(t\rightarrow \infty)$ region, so that we can directly use
\begin{equation}
\Disc F_1(t\gg m^2) \simeq {2 i\pi\beta_X \over g} \,,
\end{equation}
see Eq.~\eqref{betalimitF1}, which nicely links the contribution to the time delay of the form factors to the $\beta$-function.  This approximation is valid for scalars $\beta_{\phi}=\frac{1}{3} \left(\frac{g^3}{16\pi^2}\right) $ and fermions  $\beta_{\psi}=\frac{4}{3} \left(\frac{g^3}{16\pi^2}\right)$. The vector case is characterized by the presence of soft logs that can be resummed and therefore needs a different treatment (see \sref{sudakov}).

For  $b\ll 1/m$, we can cut the integral at some  $y = 2\bar y \lesssim O(1)$, obtaining then
\begin{align}
\hat F_1(b^2\ll 1/m^2)\simeq& -{1\over 2\pi }\log{b/b_{\mathrm{IR}}}   -\frac{\beta_X}{2\pi g}\log^2 \left(bm/\bar y\right)   \ .
\end{align}
Notice that the sign of the quantum corrections is always negative for any value of $b$ and $\bar y$. This will play a major role in the discussion about causality in \sref{causality}.

For the helicity flipping contribution, by using the dispersive representation of $F_3(0)$ (see Eq.~\eqref{discAllFs}) we can write
\begin{align}
\hat F_3^{\prime\prime}(b^2)=\frac{i}{8\pi^2 b^4}\int_{2mb}^{+\infty} dy\,  \Disc{F}_3(y^2/b^2m^2)\left[yK_2\left(y\right)-\frac{2}{y}\right] \,,
\end{align}
which gets the most important contribution from the region $t \gg m^2$ where the discontinuity converges to a delta function, $\Disc{F}_3(t \gg m^2) = i \alpha \kappa_X \delta(t)\,$ for some constant $\kappa_X $ (see \tref{disclim}). 
Therefore, we get
\begin{align}
\label{F3alternative2d}
\hat F_3^{\prime\prime}(b^2)\simeq-\frac{\alpha \kappa_X}{16\pi^2 b^2}\lim_{y\rightarrow 0}\left[K_2(y)-\frac{2}{y^2}\right] = \frac{\alpha \kappa_X}{32\pi^2 b^2} \ . 
\end{align}

 \begin{figure}[t!]
\centering \hspace{-.2in}
\includegraphics[height=5.5cm]{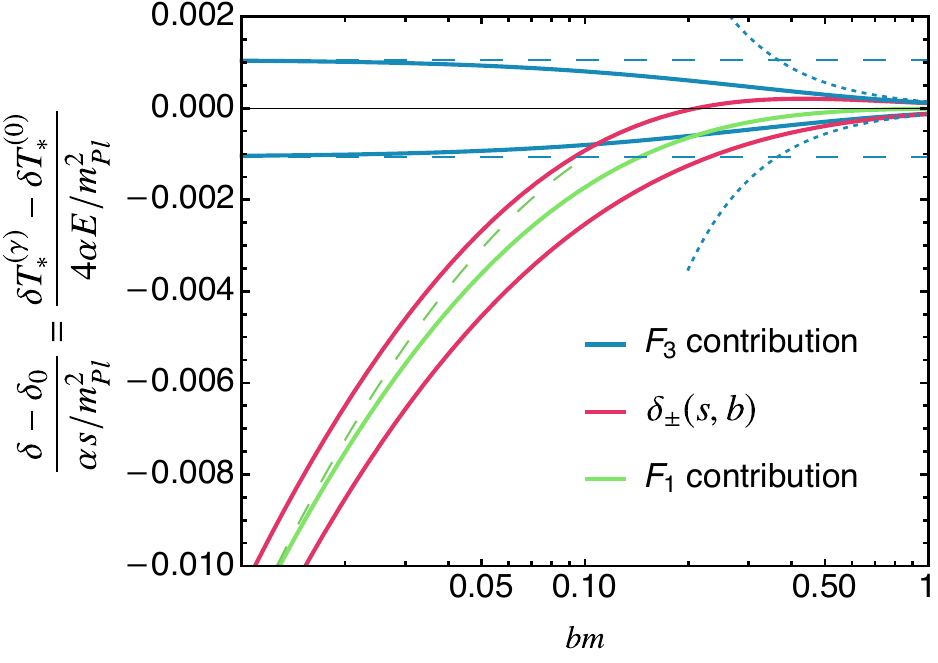} \hspace{.2in}
\includegraphics[height=5.56cm]{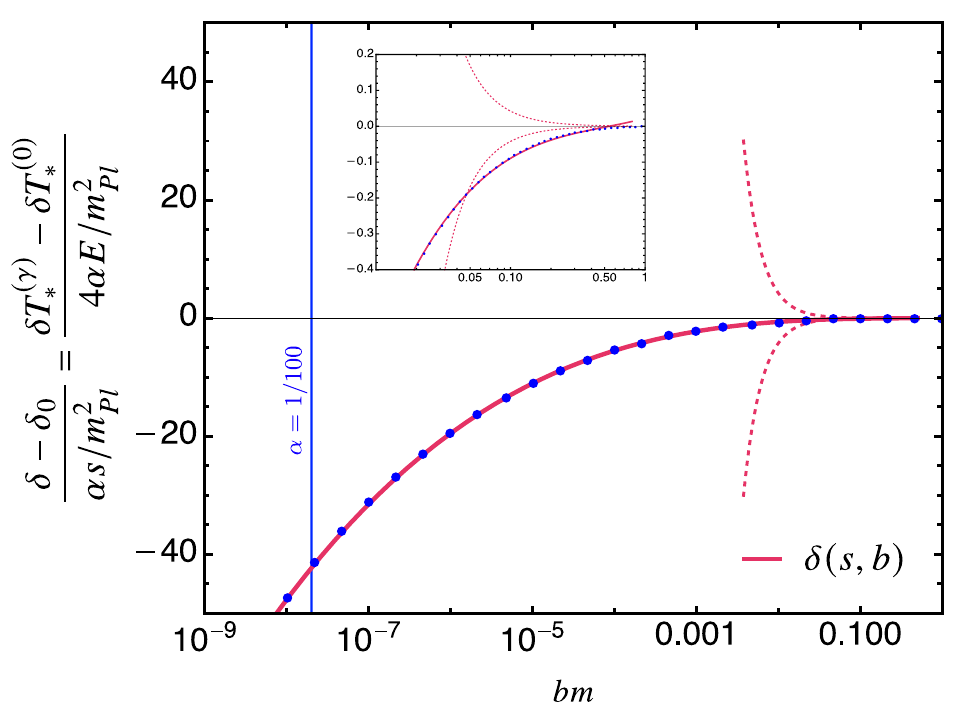}
\caption{\footnotesize
 Quantum corrections to the phase shift as function of $b m$. Dotted lines for large $b m$ show the EFT result which, if allowed to continue to small $bm$, would eventually give a negative total phase shift and thus time delay.  See discussion in \sref{asymptoticsec}. \textbf{Left:} We plot the scalar case (as discussed, the spinorial case has similar features, so it is not shown here) that have two contributions coming from the form factors $F_3(t)$, relevant at large impact parameters \eref{deltalargeblimit2}, and $F_1(t)$, which dominates as small $bm$ \eref{deltasmallblimit}. The full numerical solutions \eref{eigenvaluesPhaseShift} is shown as solid lines, and their limiting behaviors as dashed lines.  We have taken $\bar y=0.27$ to make the approximation close to the exact answer on the scales shown in the plot. \textbf{Right:} We plot the vector-loop case, i.e. QED embedded in a non-abelian gauge theory. The form factors are exponentially suppressed by  Sudakov resummation in the region $b m \ll \mathrm{exp}(-\sqrt{\pi/2\alpha})$, as discussed under \eref{sudakF1larget}, but this effect is not displayed here.   The vertical blue line represents, for $\alpha=1/100$, the value of $bm$ below which Sudakov resummation can no longer be neglected. For larger values of the impact parameter, $\mathrm{exp}(-\sqrt{\pi/2\alpha}) \ll b m \ll 1$, the fixed-order 1-loop approximation is instead accurate without resummation.  In this region, $F_1(t)$ gives the leading contribution to the phase shift \eref{deltasmallblimitWNoSudakov} and it is plotted as a solid red line which interpolates the blue dots representing the exact numerical solution.  We have taken $\bar{y}\simeq  1.12$ and $\gamma \simeq 0.12$ in \eref{approximationF1}.  \label{plots}}
\end{figure}

The full phase shift is then given by 
\bea
\label{deltasmallblimit}
\delta_\pm (  s , b \ll 1/m) =\delta_{0}(s,b) -\frac{s\beta_X}{8\pi g\mpl^2}\log^2 bm/\bar y\pm\alpha {s\kappa_X  \over 8\pi^2 \mpl^2}\,,
\eea
where $\kappa_\phi  = -1/12$ for a scalar in the loop and $\kappa_\psi =1/6$ for a fermion.
In particular, for small enough impact parameter, the $\log$ correction proportional to the $\beta$-function will dominate over the constant contribution of $F_3(t)$, as shown numerically in \fref{plots}. 

Notice, that the change in behavior of the $F_3(t)$ contribution at small impact parameter, from $1/(mb)^2$ to a constant in $b$, is crucial in the causality discussion. If that was not the case, we would observe causality violation even for the asymptotic definition (see \sref{asymptoticsec}). This is avoided thanks to the onset of new physics associated to the particles of mass $m$ before such a violation would become resolvable.   We discuss the consequences in \sref{causality}.

The inclusion of more species is straightforward, with  $\log^2 b m$ term in \eqref{deltasmallblimit} being just  replaced by the appropriate masses and rescaled by the squared charges $q_i^2$, e.g.  adding charged spin-1/2 fermions results in $\beta_{\psi}\log^2 b m_\psi  \rightarrow \beta_{\psi}\sum_i q_i^2 \log^2 b m_i $, and analogously for charged bosons. In this way, and for $b$ smaller than the top-quark scale $1/m_{\mathrm{t}}$,  one can easily include the full Standard Model contribution.

\subsubsection{The Small $b$ Limit for Vector-loops}

The small $b$ region for vector-loops is  in principle more delicate because of the IR Sudakov double-logs.  However, in the region of small impact parameter where the resummation of the double-logs is not yet important, i.e. for $ \mathrm{exp}(-\sqrt{2\pi/\alpha})\ll b^2 m^2 \ll 1$, we can still work with just the fixed-order 1-loop expressions for the form factors. 

Let's focus first on the contribution from $\Disc{F}_1$ in \eref{polediscv3} and \eref{deltacombined} by considering  the integral 
\begin{equation}
I(b^2) \equiv \frac{i}{(2\pi)^2}\int_{4m^2}^{+ \infty} dt  \frac{\Disc{F}_1(t)}{t}  K_0\left(b\sqrt{t}\right) 
\end{equation}
which can be more easily determined, up to some integration constants, by integrating its second derivative $I''(b^2)$ 
\begin{equation}
\label{2derivativeF1}
I''(b^2)  = \frac{i}{(2\pi)^2 2b^4}\int_{2mb}^{+ \infty} dy\, y K_2\left(y\right)  \Disc{F}_1\left(y^2/b^2\right) \,,
\end{equation}
where we have changed variable $y=b\sqrt{t}$. 
For small $bm$, we can cut the integral at some $y = \sqrt{e}\,\bar y \lesssim O(1)$  and approximate the Bessel function as $K_2\left(y\right)  \sim 2/y$. After performing the integral in \eref{2derivativeF1} and integrating back, we get
\begin{equation}
\label{approximationF1}
I( e^{-\sqrt{2\pi/\alpha}}\ll b^2 m^2 \ll 1) \simeq \alpha \gamma +\frac{\alpha
  }{48 \pi ^2}\left(137-4 \pi ^2\right) \log \left(bm/\bar{y}\right)-\frac{\beta_W}{2\pi g}\log
   ^2\left(bm /\bar{y} \right)+\frac{\alpha}{3 \pi ^2}\log ^3\left(bm /\bar{y} \right)
\end{equation}
where $\gamma$ is an integration constant and we recall $\beta_W = -7 g^3/16\pi^2=-7 g\alpha/4\pi$. The values of $\gamma$ and $\bar{y}$ can be estimated by fitting the numeric solution of $I(b^2)$ for small values of $b$, see Fig.~\ref{plots}.

The contribution from $\Disc{F}_3$ to $\delta_{\pm}$ is more easily calculated from \eref{eigenvaluesPhaseShift} following the same steps of the previous subsection, see \eref{F3alternative2d}, where we can use the asymptotic expression  $\Disc{F}_3(t \gg m^2) = i \alpha \kappa_X \delta(t)\,$ with  $\kappa_W=-1/4$. 
Therefore, for $\mathrm{exp}(-\sqrt{2\pi/\alpha})\ll b^2 m^2 \ll 1$ the phase shift can be approximated by 
\bea
\label{deltasmallblimitWNoSudakov}
\delta (s , b)\simeq \delta_{0}(s,b) +\frac{s}{4\mpl^2 }I(b^2) \pm\alpha {s\kappa_W  \over 8\pi^2 \mpl^2}
\eea 
where $I(b)$ is approximated by \eref{approximationF1}. Actually, the last term in \eref{deltasmallblimitWNoSudakov} can safely be dropped because it is very much subleading to the second term. 

For even smaller impact parameter,  $bm \ll \mathrm{Exp}(-\sqrt{\pi/2\alpha} )$, the IR Sudakov double-logs become large and require a resummation as performed  in section \sref{sudakov}. In this regime, we can numerically compute the contribution of $F_1(t)$ to the phase shift by exponentiating the the double-logs, that is by taking $F_1^\text{1loop}(t)\rightarrow F_1^\text{Sudakov}(t)=e^{F_1^\text{1loop}(t)-1}$ under the integral in \eref{Fihat} .  This expression for $F_1^\text{Sudakov}(t)$ is a very good approximation of the exact form factor in the two asymptotic regions of $bm$. Moreover, the contribution from $F_3$ is always subleading in this region at  small $\alpha$. 

 \begin{figure}[t!]
\centering \hspace{-.2in}
\includegraphics[height=6cm]{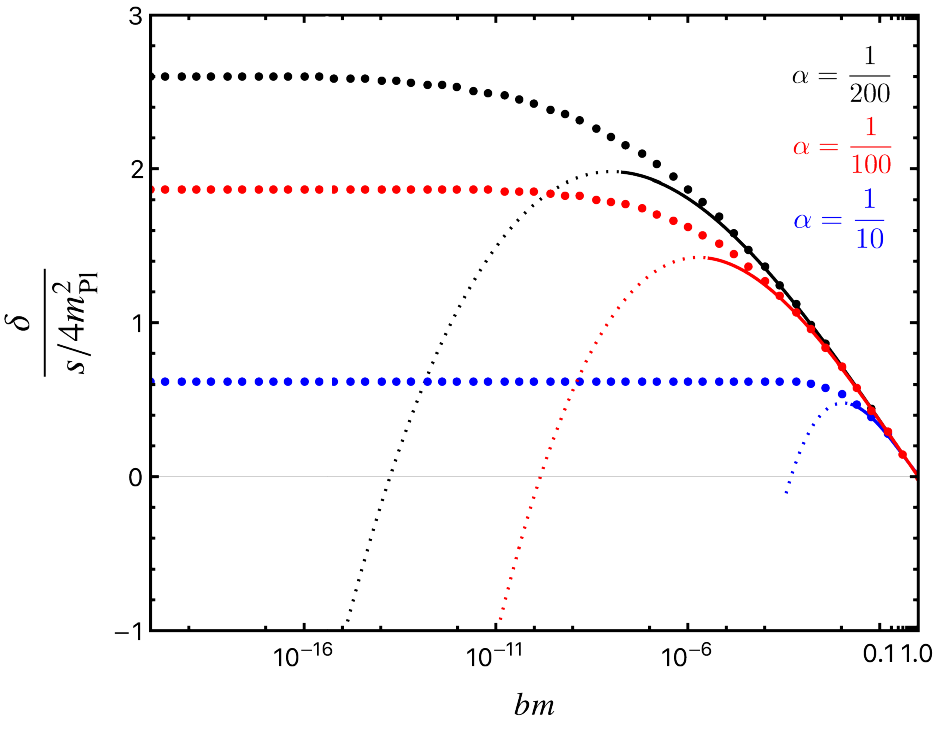}
\caption{\footnotesize
Quantum corrections to the phase shift in the vector case with IR Sudakov double-logs resummation as a function of $bm$, choosing for simplicity $b_{\mathrm{IR}} = 1/m$. The dots reproduce the full numerical solution for $\alpha=1/10$ (blue), $\alpha=1/100$ (red), and $\alpha=1/200$ (black). The solid lines for $\mathrm{Exp}(-\sqrt{\pi/2\alpha})< bm < 1$ are the analytic approximation based on \eref{approximationF1}. If extrapolated to the region $bm < \mathrm{Exp}(-\sqrt{\pi/2\alpha} )$, the phase shift would turn negative as displayed by the thin dotted lines. In such a regime of small $bm$ the \eref{approximationF1} is however no longer valid. After resumming the  IR Sudakov double-logs we find indeed a positive constant phase shift in the region $bm < \mathrm{Exp}(-\sqrt{\pi/2\alpha} )$, and no causality violation.
\label{phaseshiftsIRlogs}}
\end{figure}
Repeating the same steps as in \sref{compPhaseShift}, this integral can be computed numerically for different values of the gauge coupling. We find that for $bm \ll \mathrm{Exp}(-\sqrt{\pi/2\alpha} )$ the phase shift approaches a positive constant (see Fig.~\ref{phaseshiftsIRlogs}), and it never turns negative. 
The constant depends on the value of the gauge coupling, as it can be estimated by computing the contribution to the integral \eref{Fihat} for  $b^2 m^2 \ll \mathrm{Exp}(-\sqrt{2\pi/\alpha} )$, so that we can drop the $b$-dependence in the exponential without spoiling the convergence of the integral even in the region  $\vec{q}^{\, 2} \gg m^2 e^{\sqrt{2\pi/\alpha}}$. 
 Making the (crude) approximation $F_1^\text{Sudakov}(t) \sim e^{-\alpha/2\pi \log^2{(-t/m^2)}}$  for $\vec{q}^{\,2} \gtrsim m^2$, and $F^\text{Sudakov}_1\simeq 1$ for $\vec{q}^{\,2}\lesssim m^2$, the  \eref{Fihat}  reduces to a gaussian integral 
\bea
\label{F1Sudakov}
\hat F_1   (b^{2})  \simeq \frac{1}{2 \pi} \left[ \int^\infty_{0} d \log\left(\frac{|\vec{q}|}{m}\right) e^{-2\alpha/\pi \log^2{\left(|\vec{q}|/m\right)}}+ 
\frac{1}{2}\int_{b_{\mathrm{IR}}^{-2}}^{m^2} \frac{dq^2}{ q^2 }   \right] = \frac{1}{\sqrt{32\alpha}}+\frac{1}{4\pi}\log m^2 b^2_{\mathrm{IR}} \ , 
\eea
where we have cut off the IR divergence at $q=1/b_{\mathrm{IR}}$ (and of course $b_{\mathrm{IR}} m\geq 1$).  Therefore, the phase shift at exponentially small impact parameter with respect to the $W$'s Compton wavelength $1/m$ is 
\bea
\label{deltasmallblimitWSudakov}
\delta (s , bm  \ll \mathrm{exp}(-\sqrt{\pi/2\alpha})) \simeq {s\over 4 \mpl^2}\left[\frac{1}{\sqrt{32\alpha}}+\frac{1}{4\pi}\log m^2 b^2_{\mathrm{IR}}\right] \ , 
\eea 
which matches the numerics in Fig.~\ref{phaseshiftsIRlogs} pretty well at small $\alpha$, and shows that neither $\delta$ nor the time delay turn negative.

%
%

%
\section{Causality}\label{causality}

In this section we discuss two notions of causality that are both seemingly justified at the classical level,  but we show that in fact only one is respected in the quantum theory.
Both notions are expressed through the time delay that particles experience relative to the free time evolution in the eikonal scattering. 

\subsection{Time Delay}\label{timedelay}

We first generalize the time delay relation of \cite{Maiani:1997pd,Arkani-Hamed:2020blm} to the more general case of non-negligible inelasticity $\im \delta > 0$,  i.e. when other particles can be produced in the scattering.

The idea is to introduce a real parameter $\delta T$ that labels a family $ | f \rangle^{\delta T}_{\mathrm{out}} $ of time delayed 2-particle states, 
\begin{align}
 | f \rangle^{\delta T}_{\mathrm{out}} \equiv   \mathrm{Exp}( i\delta T H) | f \rangle_{\mathrm{out}}\,,\qquad  |f\rangle_{\mathrm{in/out}}= \int dE \sum_{J, \lambda}f_{J, \lambda}(E) | E, J, \lambda \rangle_{\mathrm{in/out}}
\end{align}
with normalization $_{\mathrm{in}}\langle f| f \rangle_{\mathrm{in}}=_{\mathrm{out}}\langle f| f \rangle_{\mathrm{out}}=\int \sum_{J, \lambda} dE |f_{E,J,\lambda}|^2=1$, and then search  for a $\delta T=\delta T_*$ that maximizes the transition probability
  $| {}_{\mathrm{out}\!\!\!}^{\delta T\!\!\!}~\langle f |f\rangle_{\mathrm{in}}|^2 $, 
  \begin{equation}
 {}_{\mathrm{out}\!\!\!}^{\delta T\!\!\!}~\langle f |f\rangle_{\mathrm{in}} =  \int dE \sum_{J, \lambda} |f_{J, \lambda}(E)|^2 e^{i(2\delta_{\lambda}(J,E)-E\delta T)}\,, 
 \end{equation}  
for some narrowly peaked wave-packet.  Here $E=\sqrt{s}$ is the total center of mass energy and $J$ and $\lambda$ label the basis where the 2-body partial-wave $S$-matrix is diagonal. Without spin, $J$ and $\lambda$ represent the angular momentum and possible other internal quantum numbers. With spin, $J$ is still the total angular momentum (or proportional to the impact parameter $b\simeq 2J/E$ in the eikonal limit) while $\lambda$ is a proxy for a linear combination of the helicity indices, $\lambda=\pm$ in the previous sections. 
Since $|f_{J, \pm}(E)|^2>0$, and  $\mathrm{Im}\,\delta_{\pm}(J,E) \geq 0$ by unitarity,  the transition probability for a localized wave-packet at $E\simeq E_0$  is maximised at the stationary phase 
\begin{equation}
\label{WignerTimedelay1}
\delta T_* =2\frac{\partial\mathrm{Re}\,\delta_{\pm}(J,E)}{\partial E}\,, \qquad  |{}_{\mathrm{out}\!\!\!\!}^{\delta T_*\!\!\!\!}~\langle f |f\rangle_{\mathrm{in}}|\simeq e^{-2\mathrm{Im}\,\delta_{\pm}(J,E_0)}
\end{equation}
where the transition amplitude is less than unity due to the opening of inelastic channels that deplete the elastic amplitude.   Notice that  \eref{WignerTimedelay1} reduces to  Wigner's formula $\delta T_* =2\frac{\partial\delta_{\pm}}{\partial E}$ when the phase shift is real. 
 In the eikonal limit of large angular momentum we can replace $\delta_{\pm}(J, E)=\delta_{\pm}(s,b)$, 
 and the notion of time delay becomes 
 \begin{equation}
 \label{eq:timedelay}
\delta T_* =2\frac{\partial\mathrm{Re}\,\delta_{\pm}(s,b)}{\partial E} 
\end{equation}
which again reduces to $\delta T_* =2\frac{\partial\delta_{\pm}(s,b)}{\partial E}$ for real $\delta_{\pm}(s,b)$.  
 Since $\delta_{\pm}(s,b)$ can be calculated in terms of the discontinuities of the form factors that at one loop are given by products of real tree-level vertices only, any imaginary part can only arise from two loops onward (to the leading post-Minkowskian order). The 1-loop phase shift is therefore real, and we can safely omit the ``$\mathrm{Re}$'' in the expression for the time delay to this order.  
In the following we sometimes leave the index $\pm$ of the phase shift implicit. 
 
In the eikonal limit the $| f \rangle^{\delta T}_{\mathrm{out}}$ can be visualised as the family of asymptotic outgoing trajectories emerging from the scattering region and specified by ---in addition to the impact parameter and the energy which are preserved in the scattering---  the time-shift $\delta T$ relative to the incoming asymptotic trajectory. The $\delta T=\delta T_*$ corresponds to to the emergent semiclassical asymptotic trajectory selected by quantum constructive interference.   

The way we derived the time delay makes clear that the asymptotic causality condition  $\delta T_*\geq 0$ that we discuss next  is meaningful only for $\delta T_*$ much larger than the quantum mechanical uncertainty $\delta T_{\mathrm{q.m.}}\sim \hbar/E$, which for the $\delta_{\pm}(s,b)\propto s$ in our gravitational setup  is just the requirement of large scattering phase shift. 

 Despite such a large phase shift, it is well known that $\delta$ remains reliably calculable in the eikonal scattering at large impact parameter in the transplanckian regime, or  
against a coherent state of spectators. The latter is nicely explained in detail in e.g. \cite{Camanho:2014apa}: the fact that the phase shift grows (at least) linearly with $s$, which is the case for e.g. Eqs.~(\ref{deltasmallblimit}), (\ref{deltasmallblimitWNoSudakov}), and (\ref{deltalargeblimit2}), implies that perturbatively small phases $\delta\ll 1$ of photons scattering against a series of time-separated $N\gg 1$ spectator particles exponentiate thanks to factorization,  $(1+i\delta)^N\rightarrow e^{iN \delta}$,  while keeping the wave-packet localized in the impact parameter space.\footnote{For the opposite regime where factorization does not hold because the spectators are on top of each other, see the interesting \cite{Kologlu:2019bco}.} Therefore, the sequence of scattering events returns $\tilde{\delta}=N\delta\gg 1$, i.e. a large scattering phase produced by a coherent state of spectators. Since $\tilde{\delta}$ and $\delta$ have the same dependence on energy and impact parameter, in the following we keep referring to just $\delta$, but it is left understood that we actually consider scattering against a coherent state that gives rise to a large phase shift.  

 Alternatively, we can resort to the reliable exponentiation of the large eikonal phase when scattering against a particle in the transplanckian regime $s\gg \mpl^2$ at small momentum exchanged $t\ll s$, see e.g. \cite{Amati:1992zb,Amati:1987uf,Kabat:1992tb} and references therein, where the theory admits as well a semiclassical approximation for impact parameters much larger than the Schwarzschild radius $R_s\sim G\sqrt{s}$.  Higher post-Minkowskian corrections correspond to including higher relative $o(R_s/b)$  corrections to $\delta$, and they  are made smaller than the gauge loop contributions  we calculated by a suitable choice of the kinematics within the transplankian eikonal scattering.   Post-Minkowskian corrections are instead clearly more important than gauge corrections when scattering astrophysical bodies.
 
Transplanckian scattering can be interpreted, to lowest order, as 1-to-1 scattering in a gravitational Aichelburg-Sexl shock-wave background at large impact parameter. Our next-to leading order analysis treats gravity as fully dynamical and away from the probe limit,  not just as in a QFT  at fixed external background that would otherwise break spacetime symmetries (as e.g. it is the case for QED in a rigid shockwave background). An effective metric  can nevertheless be associated to the scattering by amending it   order by order. The $O(G^3)$-corrected shockwave metric that gives rise to the same scattering angle as obtained from the scattering amplitudes in pure gravity has been calculated e.g. in \cite{Ciafaloni:2014esa,Amati:1993tb}.  It would be interesting to calculate the $O(\alpha)$ correction to Aichelburg-Sexl shock-wave.    The connection between scattering in a non-trivial background and causality is reminiscent of the positivity conditions for scattering amplitudes of \cite{Adams:2006sv}, and its connection to the positivity of time delay has been made explicit in \cite{Arkani-Hamed:2020blm}.

\subsection{Asymptotic Causality} \label{asymptoticsec}

By asymptotic causality we mean the following refinement of the Gao-Wald condition \cite{Gao:2000ga} used in \cite{Camanho:2014apa}:  the time delay experienced by any particle scattering against a coherent source of spectators, or in the eikonal transplankian regime,  should be positive 
\begin{equation}
 \label{CausalityCond}
\delta T_* \geq 0
\end{equation}
for all $b < b_{\mathrm{IR}}$, whenever resolvable and calculable within the range of validity of the theory.\footnote{In fact, since $\delta(b,s)$ in $D=4$ is defined only up to an overal shift $\delta(b,s)\rightarrow \delta(b,s)+{s\over 8\pi \mpl^2}\log{\lambda}$ (associated to rescaling of the IR cutoff), which results in a $b$-independent shift of the time delay $\delta T_*\rightarrow \delta T_*+ \mathrm{const}$, a slight refinement is demanding that $\delta T_*$ should be bounded below as $b$ is decreased, or simply $\lim_{b\rightarrow 0}\delta T_* > -\infty$, a condition insensitive to the rescaling of the IR cutoff.}

The violation of asymptotic causality would imply that signals sent via massless particles through the bulk of a spacetime perturbed by  some spectator field (which could be the graviton itself), would be recorded by a detector at future null infinity at earlier times than if sent instead through an unperturbed empty Minkowski spacetime, provided the impact parameter is chosen small enough.  
Notice that any violation of causality would be associated to small regions of spacetime, far from the IR cutoff, $b\ll b_{\mathrm{IR}}$, and it is relative to the flat Minkowski causal structure that is obtained by removing the massless spectator field, e.g. by sending the center of mass energy to zero. 

Turning the asymptotic causality condition around,  forbidding its violation  can be used to determine the validity range of the theory, that is putting bounds on the cutoff and/or couplings. For instance, \cite{Camanho:2014apa} put bounds on the cutoff associated to certain EFTs under the assumption that the higher-dimensional operators are generated at tree level so that the resolution of apparent causality violation should also be resolved at tree level,  as it happens in string theory that provides infinitely many higher spin states exchanged at tree-level \cite{Camanho:2014apa,DAppollonio:2015fly} to fix the issue with causality.  Tree-level causality bounds are obtained along similar reasoning in e.g. \cite{Hinterbichler:2017qyt,Bonifacio:2017nnt,Afkhami-Jeddi:2018apj, Kaplan:2019soo} and several other works.  In this paper we are instead interested in probing the notion of asymptotic causality quantum mechanically, and within QFT.   

The results of the previous sections show that this asymptotic notion of causality is in fact respected at one loop in gauge theories that are perturbatively renormalizable (before turning on gravity) at all scales. 
 The reason lies in the change of behavior of the contribution to the phase shift of $F_3(t)$, which transitions from the {\it unbounded} $1/b^2$ in the EFT regime, where charged particles are integrated out, to a constant (see \fref{plots} and Eq.~\eqref{deltasmallblimit}), without ever becoming of the size of the leading effect. Moreover, while the $F_1(t)$ contribution to the time delay $\delta T^{(F_1)}_*=- [ 4E \beta /(8\pi g \mpl^2) ] \log^2b m$ from \eref{deltasmallblimit} and \eref{eq:timedelay} {\it does} become indefinitely more negative as the impact parameter is decreased, the condition $\delta T_* \geq 0$ in fact remains  always satisfied  as long as  the impact parameter $b$ is taken larger than the strong coupling scale $b_L$ of the Landau pole (if any)\footnote{Notice that we have also taken $m b_{\mathrm{IR}}\gg 1$ while respecting $b_L b_{\mathrm{IR}}m^2\ll 1$, that is $b_{\mathrm{IR}}<1/m \, \mathrm{Exp}(g/\beta)$ which is exponentially larger than $1/m$, hence a valid choice for the IR cutoff, for perturbative couplings.  Larger values of $b_{\mathrm{IR}}$ are certainly valid, but there is no choice for which a violation of $\delta T_*>0$ can be found in the domain $b_L< b < b_{\mathrm{IR}}$. Alternatively, one can remove any $b_{\mathrm{IR}}$-dependence by  looking for the scale  $b_*$ where gravity would become repulsive, that is where the scattering angle would change sign and the photon would be deflected away,  as proxy for the scale of causality violation. This corresponds to demand gravity always being attractive. The $b_*$ has the same parametric dependence on $m$ and $g$ than  $b_L$ in \eref{landaupolebL}.  }
\begin{equation}
\label{landaupolebL}
b > b_L=\frac{1}{m}e^{-g/\beta}\qquad \mbox{for  } g/\beta>0\,.
\end{equation}
Here $\beta>0$ is the $\beta$-function of the gauge coupling  $g$ of any spin-0 and spin-1/2 charged particles running in the loop.
 
As for the quantum corrections generated by spin-1 particles, they  satisfy automatically the causality condition $\delta T_*\geq 0$ for any $b<b_{\mathrm{IR}}$  since the Sudakov IR double-logs suppress exponentially the form factor at  the same scale where quantum corrections would otherwise start dominating over the minimal gravitational contribution, see Fig.~\ref{phaseshiftsIRlogs}.   This is nicely consistent with the fact that the non-abelian gauge theories associated to  charged spin-1 particles have negative $\beta$-functions and are therefore asymptotically free in the UV,  needing  a priori no UV completion before meeting the Planck length  $1/\mpl$.    
 
 In other words, no asymptotic-causality violation is therefore detectable, even at the quantum level, at any length scale within the range of validity of our perturbative calculations.  Moreover,  because of the connection we have established between the sign of the $\beta$-function and the sign of the leading quantum corrections to the phase shift $\delta(s,b)$ at small impact parameter, demanding that $\delta T_* \geq 0$ correctly infers the existence of new dynamics at or before the scale of the  Landau pole\footnote{We are tacitly considering the case where  the scale of the Landau pole of the $U(1)$ gauge theory at hand is  smaller than the Planck mass.  For theories with $\alpha$ too small, the Landau pole would be found beyond the Planck mass and  the bound would trivialize to $b>1/\mpl$, where gravity becomes already strongly coupled. } $\Lambda_L=1/b_L$, if any. That is, in scalar and spinorial QED either new physics in the form of strong coupling or weakly coupled particles must appear at $b>\mathrm{max}\{b_L, 1/\mpl\}$, while for QED embedded in a non-abelian gauge theory with negative $\beta$-function the only consistency threshold associated  to asymptotic causality is set by the Planck length. 
 
Our finding shows that asymptotic causality is therefore able to diagnose the presence of a cutoff not only when the theory has strongly irrelevant operators like in \cite{Camanho:2014apa}, but even when the cutoff is exponentially large because it is associated with marginally-irrelevant deformations such as the gauge coupling in  QED.

\subsection{Bulk Causality}

Let's turn now to another notion of causality which stems from the idea to race against gravitons through a spacetime perturbed by  some spectator field. The bulk {(or local)} causality condition  is the statement that any massless particle would lose the race to the graviton by an amount that is resolvable and calculable within the range of validity of the theory. 
  That is, sending a massless particle and a graviton with the same energy simultaneously  through the bulk of a weakly perturbed spacetime, a detector at future null infinity would always record the graviton first and then the other particle.  

Sending photons for definiteness,  bulk causality implies 
\begin{equation}
\label{bulkcausality}
\delta T_*^{(\gamma)}  -\delta T_*^{(0)} \geq 0 \,,
\end{equation}
where $\delta T_*^{(0)}=2\partial\delta_{0}(s,b)/\partial E$ is the time delay experienced by gravitons, which is the classic Shapiro time delay.  At the classical level the difference in the time delay vanishes, i.e.  massless particles travel along the same geodesic, classically, for large impact parameter. 
The difference in \eref{bulkcausality} removes the universal term which is also present in the photon time delay  as a manifestation of the classical equivalence principle. 
Therefore, bulk causality  \eref{bulkcausality} is genuinely sensitive to quantum corrections generated by charged states running in the loop.

As a matter of fact, the quantum corrections we calculated in the previous sections violate the bulk-causality condition quantum mechanically, within the range of validity of perturbation theory. 
Indeed, at small impact parameter for loops of spin-1/2 ($X = \psi$) and spin-0 ($X=\phi$) particles, $b\ll 1/m$, the difference in time delays is 
\begin{equation}
\left(\delta T_*^{(\gamma)}  -\delta T_*^{(0)}\right)_{X} \simeq -\frac{E\beta_X}{2\pi g\mpl^2}\log^2 bm/\bar y \end{equation} 
and it is negative even for impact parameters much larger than the Landau pole length-scale $b_L$, see \fref{plots}.   
For spin-1 particles running in the loop ($X=W$), the {leading} contribution for $b^2 m^2 \ll \mathrm{Exp}(-\sqrt{2\pi/\alpha} )$ is given by \eref{deltasmallblimitWSudakov} and therefore
\begin{equation}
\label{eq:deltaTspin1}
\left(\delta T_*^{(\gamma)}  -\delta T_*^{(0)}\right)_{W} \simeq \frac{E}{\mpl^2}\left(\frac{1}{2\pi}\log(b m)+\frac{1}{\sqrt{32\alpha}}\right) \ , 
\end{equation} 
which is also negative in this regime of small impact parameter. Note that these differences are independent of the IR cut-off $b_\text{IR}$.

  All in all, bulk causality is violated at one loop:\footnote{A word of caution: one could try to restore it by adding more degrees of freedom that,  however, should be relatively light, with a mass $M$ that is at best 1-loop factor away from the charged states we considered,  in order not to decouple again their contribution to $\delta$ at the rate  $1/b^2M^2$. } although light that scatters against spectator particles   is always slower than free gravitons in  Minkowski spacetime (asymptotic causality), it can win the race against gravitons that also bounce off the same spectators.\footnote{This made the 2020/21 Tokyo Olympics much more challenging for light.}  We interpret this result as evidence against bulk causality, whereas asymptotic causality is respected at one loop. 
 
We emphasize that this conclusion is similar in spirit to the Drummond-Hathrell ``paradox'' \cite{Drummond:1979pp} where one is working directly with the velocity of the perturbations in certain backgrounds and charged states have been integrated out. In that case, however, the alleged violation of causality is not resolvable within the validity range of the EFT, see e.g. \cite{Goon:2016une,deRham:2020zyh},  whereas in our case bulk causality fails within the validity of the perturbative QED theory with dynamical gravity (as opposed to fixed background), where the propagating charged particles remain in the spectrum and where the effect is resolvable as soon as $s\gg \mpl^2 4\pi/\alpha$, that is at transplanckian energy (or scattering against several spectators).    

We remark also that our analysis is entirely performed within perturbation theory of a renormalizable gauge theory minimally (or conformally) coupled to gravity, i.e. with  small coupling $g$,  large impact parameter  $b  >  \mathrm{max}\{b_L,1/\mpl\} $ but possibly $b\ll 1/m$, where $m$ is the mass of charged states,  taking $\delta\gg1$ and small scattering angle $\theta^2\ll 1$, from either eikonal scattering against a coherent state or in the transplankian eikonal regime to the leading post-Minkowskian order. 
We  have nothing to say about scattering photons outside this domain, in contrast to e.g. \cite{Hollowood:2015elj,Hollowood:2016ryc} that find no bulk-causality violation but, as far as we understand, in a different regime where photons propagate as a probe in a  fixed shockwave background which break null-coordinate translations, generate $\mpl$- and $s$-independent corrections to $\delta-\delta_0$, and give rise to $\mathrm{Im}\delta\neq 0$ already at one gauge loop.  A QFT in a fixed curved spacetime cannot be sensitive, by construction, to bulk-causality violation, because the front-wave velocity is at best 1.  Our study is instead performed  in a post-Minkowskian expansion with dynamical gravity and away from the probe limit, where the background is not fixed, and in fact the metric has to be reconstructed order by order, as it has been already done in the literature, e.g.  \cite{Ciafaloni:2014esa,Amati:1993tb}, for purely gravitational corrections. Different species ---photons vs. gravitons--- have different interactions 
and the effective metrics are in general different except for $b\rightarrow \infty$, in particular $b\gg 1/m_X$,  where one indeed recovers the equivalence principle as an emergent low-energy effect. 

It is presently unclear what the physical consequences of bulk causality violation would be.  It appears that no fundamental principle is violated by having two particle species that travel across a shockwave slower than in Minkowski space, despite one species being relatively faster than the other one. On a practical side, however, and taking it at face value,  our finding teaches us that bulk causality should not be used to constrain EFT coefficients,  as is instead sometime advocated in the literature, see e.g. \cite{deRham:2020zyh}.  This should be contrasted with the recent gravitational positivity bounds derived in \cite{Caron-Huot:2022ugt} which are  instead based on asymptotic causality at all scales, as in the present work.

\section{Conclusions}\label{conclusion}
Causality is a fundamental concept in classical as well as in quantum physics in flat spacetime, and it is at the core of relativistic QFT.  
Its implications ---its precise incarnation--- are however presently not fully understood in dynamical gravity, once quantum matter effects are taken into account or even when spacetime is itself subject to quantum fluctuations.

In this work we have studied causality in gravity to the first post-Minkowskian order around flat spacetime,  focusing on the leading quantum effects in a gauge theory. 
We have in particular contrasted two notions of causality ---``asymptotic'' and ``bulk'' causality--- in a gauge theory where both are respected in the classical limit, but differ quantum mechanically. 

The causal and quantum response of photons to a weak spacetime perturbation is captured by the photon energy-momentum tensor that we have calculated by exploiting 
unitarity cuts and on-shell techniques, determining the 1-loop form factors generated by charged scalars, spinors, and vectors running in the loop. In analogy to the electroweak currents, vector loops give rise to large IR Sudakov $\log$-factors in the energy-momentum tensor, that we have re-summed.  

We have studied the eikonal phase shift of photons scattering against spectator particles  in the transplanckian regime (or with several subplanckian spectators building up a large scattering phase).   In the limit of large impact parameter, the theory is well approximated by an EFT with higher dimensional operators, and we have reproduced the classic result of \cite{Camanho:2014apa} and its implication for causality.  In particular, the would-be violation of asymptotic causality in the EFT can be used to determine the shortest impact parameter where the EFT must necessarily break down, to be replaced by a new theory where microscopic degrees of freedom (the charged state and possibly Higgs bosons) are integrated-in. 

Beyond the large impact-parameter limit, and studying the time delay all the way down to the Landau-pole scale,  we have explicitly established how asymptotic causality is actually respected in the gauge theory coupled to gravity, even in the UV. 
 The helicity-flipping form factor that would have led to causality violation in the EFT changes behavior as the impact parameter becomes comparable to the Compton wavelength of the particles running in the loop. Its contribution to the phase shift transitions  to a constant value in the impact parameter $b$,  subleading to the contribution from the helicity-preserving form factor which becomes instead the leading correction to the classic Shapiro time delay. 
Moreover, we find that the sign correction to the the time delay, at small impact parameter, is  opposite to the sign of the $\beta$-function generated by the particles in the loop. 

We have found  that asymptotic causality, i.e. positivity of the time delay relative to a photon travelling in unperturbed flat spacetime, is therefore respected up to the scale of the Laudau pole (if any), where the perturbative regime breaks down and our calculation is no longer valid. Conversely, the presence of the Landau pole can be correctly inferred by demanding asymptotic causality in a gravitational scattering.   For theories with a negative $\beta$-function, we find that asymptotic causality holds up to the Planck length. 

The fate of bulk causality, that is the notion that photons should travel locally slower (or at the same speed)  than gravitons in the bulk of the same dynamical spacetime, is different. Bulk causality implies that the difference between photon and graviton time delays should be non-negative, which is respected classically.  We have found instead that  quantum-mechanically photons always display in the UV a smaller time delay than gravitons, representing a  violation of bulk causality at the quantum level. While not resolvable in the IR (i.e. at length scales longer than the charged particles wavelengths),  it is instead resolvable in the UV for sufficiently large center of mass energy. 

Looking at future directions, it would be interesting to understand the implication of bulk-causality violation or, perhaps, how to recover it and why.  In particular, it would be interesting  to include next orders in the post-Minkowskian expansion in $R_s/b$ in the photon and graviton time delays, and compare the competing contributions between the gauge and gravitational couplings. We find intriguing the possibility of turning bulk causality into a statement about the swampland versus the landscape in gravitational theories, in analogy to the Weak-Gravity Conjecture \cite{ArkaniHamed:2006dz}.  Going to higher orders in the post-Minkowskian expansion is also interesting in itself  because of extra IR divergences, other than the ones we re-summed in this work, that can arise from loops and real emissions involving gravitons, see e.g. \cite{Ciafaloni:2018uwe}.  Understanding IR divergences in gravity and extracting physical observables and well defined $S$-matrix elements is a long-standing research program which has recently become phenomenologically even more relevant because of its connection to the gravitational waves emitted in black hole and/or neutron star mergers. 

Finally, while we focused in this work on the quantum corrections  to the 3-point function and their implications for causality in gravity, there has been recent progress in understanding how causality and unitarity  are imprinted in the analytic structure and positivity of 4-point functions \cite{Caron-Huot:2021rmr,Bern:2021ppb}. It would be extremely interesting to study causality via 4-point functions including the IR quantum effects from loops of massless states.  

%
%
%
%
\subsection*{Acknowledgments}
We thank Simon Caron-Huot, Pierre Vanhove, Gabriele Veneziano, Herman Verlinde, Filippo Vernizzi, Andrea Wulzer and Sasha Zhiboedov for useful discussions. We thank Riccardo Rattazzi for useful discussions and for pointing out the correct scaling in $\alpha$ in \eref{deltasmallblimitWSudakov}.  We thank Javi Serra for the continuous discussions, insights, and advice throughout this work, and his early participation.  F.S. is supported by a Klarman Fellowship at Cornell University, and also supported in part by the NSF grant  PHY-2014071.   M.L. acknowledges the Northwestern University Amplitudes and Insight group, Department of Physics and Astronomy, and Weinberg College, and is also supported by the DOE under contract DE-SC0021485. G.I. thanks the TH-department of CERN for the kind hospitality during the completion of this work.

\appendix

\section{Conventions} \label{conventions}
We are working in the mostly minus signature convention $\eta_{\mu\nu}=\mathrm{diag}(+,-,-,-)$,  with Riemann tensor $R^\mu_{\,\,\nu\rho\sigma}=\partial_\rho \Gamma^{\mu}_{\sigma\nu}+\ldots$,  Ricci tensor $R_{\nu\sigma}=R^\mu_{\,\,\nu\mu\sigma}$, Weyl tensor $W^{\mu\nu\rho\sigma}=R^{\mu\nu\rho\sigma}-\mbox{traces}$. Symmetrization and antisimmetrization does not involve factorials, e.g.  $A_{[ab]}=A_{ab}-A_{ba}$ and $A_{(ab)}=A_{ab}+A_{ba}$. The discontinuty across the real line of an analytic function $F(t)$ in the cut $t$-plane is defined as $\Disc F(t)=F(t+i\epsilon)-F(t-i\epsilon)$. 
The one particle states are normalized relativistically, $\langle k^{\prime h^\prime} |k^h\rangle=\delta^{h h^\prime}(2\pi)^{3}\delta^{(3)}(\vec{k}-\vec{k}^\prime)2\sqrt{\vec{k}^2+m^2}$, {where $\vec{k}$ is a three-dimensional vector.}

We work with the following spinor conventions. The momentum $k_i$ of a massless particle $i$ is rewritten as 
\beq
\label{SHF_lambdadef}
(k_i \sigma)_{\alpha \dot \beta} =  \lambda_{i \, \alpha} \tilde \lambda_{i \, \dot \beta},  
\eeq 
where
$[\sigma^\mu]_{\alpha \dot \beta}  = (1, \vec  \sigma )$. The objets $\lambda_i$ and $\tilde\lambda_i$ are respectively holomorphic and anti-holomorphic spinors, transforming as $\lambda_i \to t_i^{-1} \lambda_i$, $\tilde \lambda_i \to t_i \tilde \lambda_i$ under a $U(1)$ little group transformation. The spinor-helicity angle (square) brackets with positive energy correspond to negative (positive) 1/2-helicities.

A similar construction is done for massive momenta \cite{Arkani-Hamed:2017jhn}
\beq
\label{SHF_chidef}
(k_i  \sigma)_{\alpha \dot \beta} =  \sum_{J=1}^2  \chi_{i \, \alpha}^{\, J} \tilde \chi_{i \, \dot \beta \, J}\ , 
\eeq 
where the $J=1,2$ index is associated with the $SU(2)$ little group, with $\epsilon^{IJ}\chi_J=\chi^I$. For each external state, a full symmetrization of these indices is necessary to reproduce different polarizations. For clarity, the $J$ index as well as the symmetrization is kept implicit in the main text.

Lorentz-invariant spinor contractions are abbreviated using the bra-ket notation.
For massless spinors 
\beq
 \langle i j \rangle \equiv \lambda_i^\alpha \lambda_{j \, \alpha} =  \eps^{\beta \alpha } \lambda_{i \, \alpha} \lambda_{j\, \beta} , 
\qquad  
  [ i j ] \equiv \tilde \lambda_{i\, \dot \alpha} \tilde \lambda^{\dot \alpha}_j   = \eps^{\dot \alpha  \dot\beta} \tilde \lambda_{i \, \dot \alpha} \lambda_{j \, \dot \beta} \, , 
\eeq  
where $\epsilon$ is the anti-symmetric tensor. We use the notation $\langle i  p_k j]\equiv \lambda_i p_k \sigma \tilde \lambda_j =\la ik\ra [kj]$.  Similarly, for massive spinors 
\beq
 \langle \mathbf{i j} \rangle \equiv  \eps^{\beta \alpha } \chi_{i \, \alpha}^J \chi_{j\, \beta}^K\ ,\qquad  
 [ \mathbf{i j}  ] \equiv 
\eps^{\dot \alpha  \dot\beta} \tilde \chi_{i \, \dot \alpha}^J \chi_{j \, \dot \beta}^K \, .
\eeq  
The {\bf bold} notation allows to differentiate between massive and massless states. Finally, the Mandelstam variables are given by
\beq
s_{ij}=(k_i+k_j)^2=\la ij\ra [ji] \ .
\eeq

%
%

\section{Free Energy-Momentum Tensors}
\label{FreeTmunu}
The free energy-momentum tensor for photons $\gamma$,  a minimally coupled massless neutral (scalar) spectators $S$, and charged spinning particles with $J\leq 1$ 
\begin{align} 
\label{photonTmunu}
T^{(\gamma)}_{\mu\nu}  = & -F_{\mu\alpha}F_{\nu}^{\,\,\alpha}+\frac{1}{4}\eta_{\mu\nu}(F_{\alpha\beta}F^{\alpha\beta}) \nonumber  \\
T^{(S)}_{\mu\nu}  = & \partial_\mu S \partial_\nu S -\frac{\eta_{\mu\nu}}{2} (\partial S)^2  \nonumber  \\ 
T^{(J=0)}_{\mu\nu}(x) =& \partial_\mu\phi^\dagger \partial_\nu \phi +\partial_\mu \phi \partial_\nu \phi^\dagger -\eta_{\mu\nu}\left( |\partial\phi|^2-m^2|\phi|^2\right) 
-\frac{\xi}{3}\left(\partial_\mu\partial_\nu-\eta_{\mu\nu}\square \right)|\phi|^2     \\
T^{(J=1/2)}_{\mu\nu}=&\frac{1}{4} \bar{\psi} \gamma_{(\mu}i \overleftrightarrow\partial_{\nu)}\psi -\eta_{\mu\nu}\bar{\psi} (i \frac{\overleftrightarrow{\slashed\partial}}{2} -m)\psi   \nonumber  \\ 
T^{(J=1)}_{\mu\nu}(x) =& -W^\dagger_{\mu\alpha}W_{\nu}^{\,\,\alpha}-W_{\mu\alpha}W_{\nu}^{\dagger \alpha}+\frac{1}{2}g_{\mu\nu}(W^\dagger_{\alpha\beta}W^{\alpha\beta})+ m^2_W \left(W_\mu^\dagger W_\nu+W_\mu W_\nu^\dagger \right) -m_W^2\eta_{\mu\nu}W^\dagger_\alpha W^\alpha \ , \nonumber
\end{align}
sum up to the total free energy-momentum tensor $T_{\mu\nu}(x)=T^{(\gamma)}_{\mu\nu}(x)  +T^{(S)}_{\mu\nu}(x)+\sum_{J\leq 1} T^{(J)}_{\mu\nu}(x)$.  Above, $W_{\mu \nu}$ is the field strength related to the spin-1 field $W_\mu$.  The scalar contribution $T^{(J=0)}_{\mu\nu}(x) $ is defined only up to an improvement term controlled by $\xi$. It can be derived from the action $\int d^4 x \sqrt{|g|}\left\{ |\partial\phi|^2 -m^2|\phi|^2 +\frac{\xi}{6}|\phi|^2 R\right\}$  via $\int d^4 x \sqrt{|g|} \frac{1}{2}T_{\mu\nu} \delta g^{\mu\nu}=\delta S$, where  $\xi=0$ corresponds to a minimally coupled scalar whereas for $\xi=1$ the scalar is conformally coupled, the action being classically scale invariant when $m=0$. At the quantum level the $T_{\mu\nu}$ does not mix with other conserved operators for $\xi=1$. For the scalar spectator field we chose $\xi=0$ for simplicity. The phase shift in the eikonal limit is insensitive to the $F_2$ form factors and it is therefore independent of $\xi$.  

\section{Higgs/Graviton Mixing}
\label{HiggsGravitymix}
In this appendix we show that \eref{redefSnonmin}, which is the non-trivial Higgs boson contribution to the scattering of photons against spectators ---neither of which na\"ively couples to the Higgs---  can be phrased in terms of  Higgs/graviton mixing. After resolving the mixing the Higgs couples to all fields with non-vanishing energy-momentum trace.  

The the 3-point vertex $-h_{\mu\nu}T^{\mu\nu}/\mpl$ between gravitons and Higgs generates indeed a kinetic mixing after shifting field around the VEV 
\begin{equation}
-h_{\mu\nu} \delta T_{\mu\nu}/\mpl= \left(\frac{v\xi_H}{3\mpl} \right)h_{\mu\nu}\left(\partial_\mu\partial_\nu-\eta_{\mu\nu}\square \right)H+\ldots 
\end{equation}
The mixing is removed at the linear order by the field redefinition (a linearized Weyl transformation)
$h_{\mu\nu}\rightarrow h_{\mu\nu}+\eta_{\mu\nu}H \left(\frac{v\xi_H}{6\mpl} \right)$  
 as one can check following the transformation of the graviton kinetic term, namely $ -2 h_{\mu\nu}\left(\partial_{\mu\nu}-\eta_{\mu\nu}\square \right)H \left(\frac{v\xi_H}{6\mpl} \right)$. 
The transformation generates as well a coupling between the Higgs and any particle the graviton was coupling to, in particular to the trace $T^{\mu(S)}_{\mu}=-(\partial S)^2$ of the massless spectators $S$, namely  
\begin{equation}
-\frac{1}{\mpl}h^{\mu\nu}T^{(S)}_{\mu\nu}\rightarrow -\frac{1}{\mpl}\left(\frac{v\xi_H}{6\mpl} \right) T^{\mu(S)}_{\mu}H
\end{equation}
so that the scattering of photons with spectators receives a contribution at 1-loop by the Higgs boson exchange as reported in \eref{redefSnonmin}.

\section{Explicit Form Factors at One Loop}
\label{FormFactorsExplicit}
In this appendix we summarize the full results of the loop calculations for $\Disc F_i(t)$ and the full form factors $F_i(t)$ integrated by the dispersion relations \eref{discAllFs}.  In the following, we set $\tau=1-4m^2/t$ for brevity. 

In the case of fermion loops, we find perfect agreement with the results of \cite{Berends:1975ah,Milton:1977je}.  Comparing our results with Ref. \cite{Coriano:2011zk,Armillis:2010qk} that discuss spin-1 and Higgs contributions  we find excellent but not perfect agreement. A small discrepancy is found for the  $F_1$ form factor: rather than  $[5m_W^2+7s]$ in front to $-1/2 D_0$ in the first line of Eq.~(149) of Ref.~\cite{Coriano:2011zk}, we have  $[5m_W^2+\frac{7}{2}s]$.  This discrepancy corresponds to a difference in the imaginary parts by an amount  $\Disc \delta F_1= \frac{7}{2} i\alpha \sqrt{1-\frac{4 m_W^2}{s_{13}}}$. It is quite remarkable that the contribution of tens of diagrams in \cite{Coriano:2011zk} is reproduced in the present work by integrating the discontinuity of just one or two diagrams. 

\subsection*{Electron loops  ($X = \psi$)}

\begin{subequations}
\label{AllFsSpinhalf}
\begin{align}
\Disc F_1(t)=&   \frac{2i \alpha }{3t^2} \left(\sqrt{ \tau } \left(5 m^2+t\right)t-6 m^2 \left(2 m^2+t\right) \tanh ^{-1}\sqrt{ \tau } \right)\theta(t-4m^2) \\
\Disc F_2(t)=& i\frac{2\alpha m^2}{t^3}  \left(t \sqrt{\tau}-4 m^2 \tanh ^{-1} \sqrt{\tau} \right) \theta(t-4m^2)  \\
\Disc F_3(t)=& i \frac{2 \alpha m^2}{t^3} \left(-3 t \sqrt{\tau}+2 \left(2 m^2+t\right) \tanh ^{-1}\sqrt{\tau}\right) \theta(t-4m^2) \,. 
\end{align}
\begin{align}
F_1(t)=&1+ \frac{\alpha}{\pi}\left[ -\frac{13 \tau }{12}+\frac{3}{16} \log ^2\left(1-\frac{2}{\sqrt{\tau }+1}\right)+\frac{37}{18}+\frac{\tau -4}{4} \tau \left(\coth ^{-1}\sqrt{\tau }\right)^2+\frac{5 \tau -9}{6} \sqrt{\tau} \coth ^{-1}\sqrt{\tau } \right]\\
F_2(t)=& \frac{\alpha}{\pi}\left[  \frac{(\tau -1)}{192 m^2} \left(36 \tau -3 \log ^2\left(1-\frac{2}{\sqrt{\tau }+1}\right)-32\right)  -\frac{\sqrt{\tau} (\tau -1)^2}{8 m^2}\coth ^{-1}\sqrt{\tau}   \right. \\ 
\nonumber & \hspace{3.5in}  \left. -\frac{(\tau -2) \tau (\tau -1)}{16 m^2}  \left(\coth ^{-1}\sqrt{\tau}\right)^2 \right]\\
 F_3(t)= & \frac{\alpha}{\pi}\Bigg\{ (\tau -1) \left(-\frac{7 \tau }{16 m^2}+\frac{3 \log ^2\left(1-\frac{2}{\sqrt{\tau }+1}\right)}{64 m^2}+\frac{11}{24 m^2}\right)+\frac{3 \sqrt{\tau } (\tau-1)^2 \coth ^{-1}\sqrt{\tau }}{8 m^2}  \\
  \nonumber & \hspace{3.5in} +  \frac{(\tau -4) \tau  (\tau-1) \left(\coth ^{-1}\sqrt{\tau }\right)^2}{16 m^2} \Bigg\}
\end{align}
\end{subequations}

\subsection*{Scalar loops ($X = \phi$)}
\begin{subequations}
\label{AllFsScalar}
\begin{align}
\Disc F_1(t)=&  \frac{i \alpha }{6 t^2} \left(t \left(t-10 m^2\right) \sqrt{ \tau }+24 m^4 \tanh ^{-1}\sqrt{ \tau }\right)\theta(t-4m^2) \\
\Disc F_2(t)=& \frac{i \alpha}{t^3}  m^2 \left(-t \sqrt{\tau} +2 \left(2 m^2+t\right) \tanh ^{-1}\sqrt{\tau} \right)\theta(t-4m^2) \\ 
\nonumber &-\frac{4 i \alpha  m^2 \xi_\phi  \tanh ^{-1}\sqrt{\tau}}{3 t^2}\theta(t-4m^2) \\
\Disc F_3(t)=& \frac{i \alpha  m^2}{t^3} \left(3 t \sqrt{\tau}-2 \left(2 m^2+t\right) \tanh ^{-1}\sqrt{\tau}\right)\theta(t-4m^2) 
\end{align}
\begin{align}
F_1 (t)=& 1+\frac{\alpha}{8 \pi  t} \left[-\frac{1}{4} t \log ^2\left(\frac{\sqrt{\tau}-1}{\sqrt{\tau}+1}\right)+\left(t-\frac{16 m^4}{t}\right) \left(\coth ^{-1}\sqrt{\tau}\right)^2   \right. \\
\nonumber
& \hspace{2.5in}  \left. -\frac{4}{3} \sqrt{\tau} \left(t-10 m^2\right) \coth ^{-1}\sqrt{\tau}-\frac{52 m^2}{3}+\frac{19 t}{9}\right] \\ 
F_2 (t)= & -\frac{5 \alpha}{24 \pi  t} \left[1+\frac{\tau \left((4 \xi_\phi -9) t-12 m^2\right)}{5 t} \left(\coth ^{-1}\sqrt{\tau}\right)^2 +\frac{1}{20} (9-4 \xi_\phi ) \log ^2\left(\frac{\sqrt{\tau}-1}{\sqrt{\tau}+1}\right) \right. \\ 
\nonumber
&  \hspace{2.5in} \left. +\frac{36 m^2}{5 t}-\frac{24 m^2 \sqrt{\tau} }{5 t}\coth ^{-1}\sqrt{\tau}-\frac{4 \xi_\phi }{5}\right]\\ 
F_3 (t)=& \frac{\alpha}{24 \pi  t}  \left[1+\left(\frac{48 m^4}{t^2}+\frac{24 m^2}{t}-9\right) \left(\coth ^{-1}\sqrt{\tau}\right)^2+\frac{84 m^2}{t} \right. \\
 \nonumber
 & \hspace{2.5in}  \left. +\frac{9}{4} \log ^2\left(\frac{\sqrt{\tau}-1}{\sqrt{\tau}+1}\right)-\frac{72 m^2 \sqrt{\tau}}{t} \coth ^{-1}\sqrt{\tau} \right]
\end{align}
\end{subequations}

\subsection*{Vector loops  ($X = W$)}

\begin{subequations}
\label{AllFsVector}
\begin{align}
\Disc F_1(t)=&  \frac{-i \alpha}{2 t^2}  \left( t \sqrt{ \tau } \left(10 m^2+7 t\right)-8 \left(m^2+t\right) \left(3 m^2+t\right) \tanh ^{-1}\sqrt{ \tau }\right)\theta(t-4m^2) \\
\Disc F_2(t)=& -\frac{i \alpha  m^2}{t^3}  \left(3 t \sqrt{\tau}+2 \left(t-6 m^2\right) \tanh ^{-1}\sqrt{\tau}\right)\theta(t-4m^2) \\ 
\Disc  F_3(t)= & - \frac{3 i \alpha m^2}{t^3}   \left( -3 t \sqrt{\tau} +2 \left(2 m^2+t\right) \tanh ^{-1}\sqrt{\tau}\right)\theta(t-4m^2) 
 \end{align}
\begin{align}
F_1(t)=& 1- \frac{\alpha}{2 \pi  t}  \left[ \frac{35}{16} t \log ^2\frac{\sqrt{\tau}-1}{\sqrt{\tau}+1}+\left(\frac{12 m^4}{t}+16 m^2-\frac{19 t}{4}\right) \left(\coth ^{-1}\sqrt{\tau}\right)^2 \right. \\ 
\nonumber &  \hspace{2in}  \left. -\sqrt{\tau} \left(10 m^2+7 t\right) \coth ^{-1}\sqrt{\tau}+13 m^2+\frac{125 t}{12}\right] \\
F_2(t)=& -\frac{\alpha} {8 \pi  t} \left[ -\frac{\tau \left(12 m^2+t\right)}{t} \left(\coth ^{-1}\sqrt{\tau}\right)^2+\frac{36 m^2}{t}+\frac{1}{4} \log ^2 \frac{\sqrt{\tau}-1}{\sqrt{\tau}+1}  \right. \\ 
\nonumber &   \hspace{2in} \left. -\frac{24 m^2 \sqrt{\tau}}{t} \coth ^{-1}\left(\sqrt{\tau}\right)-3\right] \\
F_3(t)= &- \frac{\alpha }{\pi  t}  \left[\left(-\frac{6 m^4}{t^2}-\frac{3 m^2}{t}+\frac{9}{8}\right) \left(\coth ^{-1}\sqrt{\tau}\right)^2-\frac{84 m^2+t}{8 t}-\frac{9}{32} \log ^2\frac{\sqrt{\tau}-1}{\sqrt{\tau}+1} \right. \\
\nonumber & \hspace{2in}   \left. +\frac{9 m^2 \sqrt{\tau} }{t} \coth ^{-1}\left(\sqrt{\tau}\right)\right]
 \end{align}
\end{subequations}

\subsection*{Higgs boson contribution}

The model-dependent Higgs boson contribution can be found in \eref{Higgscontribution} that for a Higgs coupled to unit-charge spin-1 boson gives \eref{F2frommixing}.

\section{Off-Shell Vertices}
\label{lagrangianInt}
All amplitudes presented in this paper can be recovered by Feynman diagrams starting from the following Lagrangians. 
For scalars, the coupling to the gauge field is given by the covariant kinetic term $|D_\mu\phi|^2$, which expanded leads to
\begin{align}
\mathcal{L}^{J=0}= & |\partial_\mu\phi|^2-m^2|\phi|^2-ig(\partial_\mu\phi^+\phi^--\phi^+\partial_\mu\phi^-)A^\mu+   g^2A_\mu A^\mu|\phi|^2 \, .
\end{align}
In the same way the Lagrangian of spinors and massive vectors coupled to the gauge field are given by
\begin{align}
\mathcal{L}^{J=1/2}= &i\bar\psi\gamma^\mu\partial_\mu\psi-m\bar\psi\psi-g\bar\psi\gamma^\mu A_\mu\psi \, ,
 \end{align}
\bea
\mathcal{L}^{J=1}= & - &\frac{1}{2}W_{\mu\nu}^\dagger W^{\mu\nu} +m_W^2 W^\dagger_\mu W^\mu \nnl 
&+ & igF_{\mu\nu} W_\mu W^\dagger_\nu  -igA_\mu\left(W_{\mu\nu}W^\dagger_\nu-W^\dagger_{\mu\nu}W_\nu \right)-g^2 \left(A_\mu^2 |W_\nu|^2-|A_\mu W^\mu|^2\right) \, .
\eea
Additionally to the trilinear coupling of photons to the $U(1)$ global current there is also a non-minimal trilinear coupling $F_{\mu\nu} W_\mu W^\dagger_\nu$ with a tuned coefficient (as set by embedding it in a non-abelian theory) required to produce the gyromagnetic factor value of 2.


\bibliography{3dscattering} 
\bibliographystyle{utphys}

\end{document}